\documentclass[a4paper,fleqn,usenatbib]{mnras}

\usepackage{mathptmx}
\usepackage[T1]{fontenc}
\usepackage{ae,aecompl}

\usepackage{graphicx}	% Including figure files
\usepackage{amsmath}	% Advanced maths commands
\usepackage{amssymb}	% Extra maths symbols
\usepackage{xspace}
\usepackage{ulem}
\usepackage{pdflscape}

\usepackage[flushleft]{threeparttable}

\newcommand{\kep}{\textit{Kepler}\xspace}
\newcommand{\ktwo}{\textit{K2}\xspace}
\newcommand{\kmag}{$K_{\rm P}$\xspace}

\title[A PSF-based approach to \kep/\ktwo data. II.]{A PSF-based
  approach to \kep/\ktwo data. II. Exoplanet candidates in Praesepe
  (M\,44)\thanks{Based on observation with the \kep telescope and with
    the Schmidt 67/92 cm telescope at the Osservatorio Astronomico di
    Asiago, which is part of the Osservatorio Astronomico di Padova,
    Istituto Nazionale di AstroFisica.}}

\author[Libralato et al.]{M.\ Libralato\thanks{E-mail:\href{mailto:mattia.libralato@unipd.it}{mattia.libralato@unipd.it}}$^{1,2}$,
  D.\ Nardiello$^{1,2}$,
  L.\ R.\ Bedin$^{2}$,
  L.\ Borsato$^{1,2}$,
  V.\ Granata$^{1,2}$,
  \newauthor
  L.\ Malavolta$^{1,2}$,
  G.\ Piotto$^{1,2}$,
  P.\ Ochner$^2$,
  A.\ Cunial$^{1,2}$,
  V.\ Nascimbeni$^{1,2}$ \\~\\
  $^{1}$ Dipartimento di Fisica e Astronomia, Universit\`a di Padova, Vicolo dell'Osservatorio 3, Padova, I-35122, Italy \\
  $^{2}$ INAF-Osservatorio Astronomico di Padova, Vicolo dell'Osservatorio 5, Padova, I-35122, Italy \\
}

\date{Accepted 2016 August 1. Received 2016 July 12; in original form 2016 June 17}

\pubyear{2016}

\begin{document}
\label{firstpage}
\pagerange{\pageref{firstpage}--\pageref{lastpage}}
\maketitle

\begin{abstract}
  In this work we keep pushing \ktwo data to a high photometric
  precision, close to that of the \kep main mission, using a
  PSF-based, neighbour-subtraction technique, which also overcome the
  dilution effects in crowded environments. We analyse the open
  cluster M\,44 (NGC~2632), observed during the \ktwo Campaign 5, and
  extract light curves of stars imaged on module 14, where most of the
  cluster lies. We present two candidate exoplanets hosted by cluster
  members and five by field stars. As a by-product of our
  investigation, we find 1680 eclipsing binaries and variable stars,
  1071 of which are new discoveries. Among them, we report the
  presence of a heartbeat binary star. Together with this work, we
  release to the community a catalogue with the variable stars and the
  candidate exoplanets found, as well as all our raw and detrended
  light curves.
\end{abstract}


%%%%%%%%%%%%%%%%%%%%%%%%%%%%%%%%%%%%%%%%%%%%%%%%%%

%%%%%%%%%%%%%%%%% BODY OF PAPER %%%%%%%%%%%%%%%%%%

\defcitealias{Lib16}{Paper~I}

%%%%%%%%
\section{Introduction}
%%%%%%%%

\ktwo mission \citep{How14} has further boosted the ``gold rush'' of
the exoplanet hunting. While many exoplanets have been found around
field stars, only a few of them have been detected orbiting around
stellar-cluster members. The discovery and characterisation of these
cluster-hosted objects is very important to add a new piece to their
puzzling formation and evolutionary scenarios.

\ktwo, as well as \kep \citep{Bor10}, gives us the opportunity to
search for exoplanet candidates for radial-velocity follow-ups in
different open (with ages spanning from about the 100 Myrs of Pleiades
to the $\sim$8.3 Gyr of NGC~6791) and globular (M\,4 and M\,80)
clusters.

In our first work, \citep[hereafter Paper~I]{Lib16}, we showed that by
using a high-angular-resolution catalogue and point-spread functions
(PSFs) we are able to pinpoint each star in the adopted input
catalogue into each \ktwo exposure and to measure its flux after all
detectable close-by neighbours are subtracted from the image. This
PSF-based technique allows us to (i) increase the number of analysable
objects in the field, (ii) estimate an unbiased flux for a given
source, (iii) extract the light curve (LC) of a star in a crowded
environment and (iv) improve the photometric precision reachable for
faint stars ($K_{\rm P} \gtrsim 15.5)$.

In this work, we continue our effort on stellar clusters, focusing on
the open cluster (OC) Praesepe (NGC~2632, hereafter simply M\,44) that
was observed between 2015 April 27 and 2015 July 10 during the \ktwo
Campaign 5 (hereafter, C5). We applied our PSF-based approach
described in \citetalias{Lib16} to extract the LCs of the stars imaged
on the isolated target-pixel files (TPFs) of module 14, where most of
the cluster stars are observed. Although M\,44 is sparser than the
field studied in \citetalias{Lib16} and not in a super-stamp, our
technique is perfectly suitable to also analyse this cluster.

M\,44 is one of the few stellar clusters in which exoplanets have been
detected. Using the radial-velocity (RV) technique, two hot Jupiters
were found by \citet{Quinn12}, each of them around a M\,44
main-sequence (MS) star. Later, in a long-term RV monitoring of M\,44
members, \citet{Mal16} found an additional, massive Jupiter in a
very-eccentric orbit hosted by one of the two aforementioned MS stars,
discovering de facto the first multi-planet system in an OC. From the
photometric point of view, before \ktwo started operations, during the
Kilodegree Extremely Little Telescope (KELT) survey \citet{Pep08}
revealed two transiting exoplanet candidates in M\,44 field, but their
proper motions exclude their membership to the cluster (see
Sect.~\ref{exolitsect}). Driven by these promising results, we
explored the \ktwo/C5 data to search for additional (transiting)
exoplanets hosted by M\,44 members, taking advantage of the
high-precision photometry and the almost-uninterrupted time series
released by \ktwo.

%%%%%%%%
\section{Data reduction}
%%%%%%%%

%%%%%%%%%%%
\subsection{Asiago Schmidt telescope}
%%%%%%%%%%%

As in \citetalias{Lib16}, we used a high-angular-resolution input list
to perform our neighbour-subtraction technique. We observed M\,44 in
seven nights (between 2015 March 27 and April 22) using the Asiago
67/92 cm Schmidt telescope on Mount Ekar. A SBIG STL-11000M camera,
equipped with a Kodak KAI-11000M detector (4050$\times$2672 pixel$^2$
with a pixel scale of 0.8625 arcsec pixel$^{-1}$), is placed at the
focus of the telescope and covers a field of view (FoV) of about
58$\times$38 arcmin$^2$.

\begin{figure*}
  \centering
  \includegraphics[width=\textwidth, keepaspectratio]{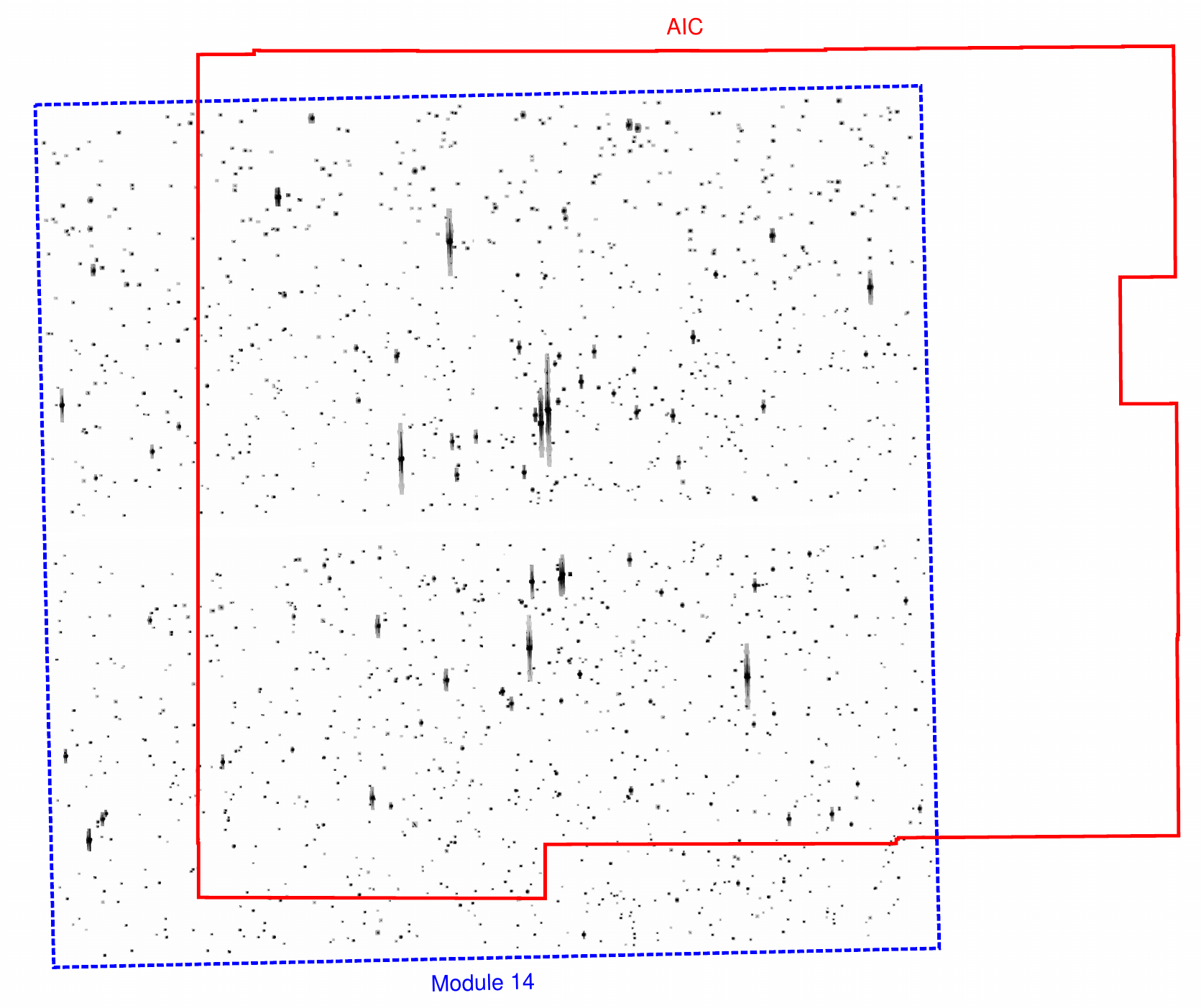}
  \caption{\ktwo module-14 FoV covered during C5. The image is a
    mosaic of four stacked images, one for each channel. Each stacked
    image was obtained by combining all 3620 usable exposures that we
    used in the LC extraction. The blue, dashed rectangle represents
    \ktwo module 14, while the red, solid rectangle shows the
    Asiago-Input-Catalog (AIC) coverage. The image is in logarithmic
    grey scale; North is up and East to the left.}
  \label{fig1}
\end{figure*}

The FoV covered by our observations is of about 2.7$\times$2.3 deg$^2$
on the sky, and was obtained by adopting a specific, dithered
observing strategy in order to cover the most of the cluster. However,
this observing campaign was performed prior to the \ktwo/C5 data
release, therefore the field overlap between our Asiago Schmidt and
the \ktwo/C5 data is not perfect (see Fig.~\ref{fig1}). In total, we
collected 120-s exposures in white light (unfiltered solution,
hereafter $N$ filter; 123 images), $B$ (48), $R$ (81) and $I$ (81)
filters. For each image, we created a set of 9$\times$5
spatially-varying, empirical PSFs and used them to measure positions
and fluxes for all the detectable objects in the field. The dedicated
software was developed starting from the work of \citet{And06} with
the wide-field imager at the 2.2\,m MPI/ESO telescope. Stellar
positions were also corrected for geometric distortion.

The input catalogue was built as described in detail by
\citet{Nar15}. Briefly, we started by making the $N$-filter input
list. We transformed (by mean of six-parameter linear transformations)
all $N$-filter stellar catalogues into the reference frame system of
the best (minimum of the product between airmass and seeing) image. We
then created a stacked image which high signal-to-noise ratio (SNR)
allowed us to better analyse faint sources. As for the single
exposures, we generated an array of spatially-varying, empirical PSFs
and measured all detectable objects over the entire FoV covered by the
stacked image. Spurious detections and PSF artefacts were removed from
the input list by using a parameter called quality of the PSF fit
\citep[\textsf{QFIT},][]{And08}, the method described in \citet{Lib14}
and by visually inspecting the stacked image.

The same procedure was performed to obtain the stellar list for the
other filters. The $B$, $R$ and $I$ magnitudes were calibrated by
using the catalogue of \citet{An07}. We selected a sample of bright,
unsaturated stars in our catalogue in common with that of
\citeauthor{An07}, and performed a least-square fit to find the
coefficients of the calibration equations. We found that a linear
relation was enough to register our photometry.

Finally we cross-identified all stars among the different catalogues
and create a multi-filter input list for M\,44. The catalogue, that
contains about 24\,000 stars measured in $N$ filter, was also linked
to the Two Micron All-Sky Survey \citep[2MASS,][]{Skru06} catalogue to
have for each star a $J_{\rm 2MASS}$-, $H_{\rm 2MASS}$-, $K_{\rm
  2MASS}$-magnitude entry (when available), and to the PPMXL catalogue
\citep{RDS10} for the ($\mu_\alpha \cos\delta$,$\mu_\delta$) proper
motions. Hereafter we refer to this catalogue as the Asiago Input
Catalogue (AIC).

%%%%%%%%%%%
\subsection{\ktwo}
%%%%%%%%%%%
\label{k2data}

The \ktwo/C5 data set was reduced following the prescriptions given in
\citetalias{Lib16}. We analysed the entire module 14 in which M\,44 is
mainly imaged, namely channels 45, 46, 47 and 48. For each channel, we
reconstructed a full-frame exposure for each of the 3620 usable
(no-evident trailing effects) \ktwo/C5 cadence number of the TPFs. The
average \kep Barycentric Julian Day (KBJD) of all the TPFs with the
same cadence number was then set as KBJD of the reconstructed
image. As in \citetalias{Lib16}, the column \texttt{FLUX} was used to
assign the pixel values.

To model the undersampled PSF of each \kep channel, we used again the
effective-PSF (ePSF) formalism of \cite{AK00}. The ePSFs were mainly
modelled following the prescriptions given in \citetalias{Lib16}. Here
we describe only the differences between the two works.

First, we used the AIC, transformed into the reference frame system of
each image, to pinpoint the position of the bright, unsaturated stars
selected to model the ePSF. The high angular resolution and the
astrometric accuracy of the AIC allowed us to better place the ePSF
samplings and overcome the most of the pixel-phase errors. Three (out
of four) module-14 channels are not completely covered by the AIC
(Fig.~\ref{fig1}). However, since our aim was to obtain a reliable,
average ePSF model for each channel, this partial coverage was a good
compromise to deal with.

Second, we introduced a neighbour-subtraction stage during the
(iterative) ePSF-modelling process. Before collecting the ePSF
samplings from a given star and model the ePSF, we subtracted (by
using the current ePSF model) all its close-by neighbours contained in
the AIC in order to decrease the light-contamination effects that
would result in a shallower ePSF. A more detailed description of the
new method will be published in a subsequent paper of this series
focused on the globular cluster M\,4 observed during \ktwo Campaign 2
(Libralato et al., in preparation), in which the improvement using
this approach is more evident due to the higher level of crowding with
respect to M\,44.

Once we converged to an average ePSF model for each channel, we
perturbed it for each image to take into account the temporal
variation. We performed a 2$\times$2 perturbation that also partially
solved for the ePSF spatial variations across the channel FoV. The
2$\times$2 array was chosen as a compromise between modelling the
spatial variations of the ePSF and having enough stars to model the
ePSF itself in each cell. We also introduced a neighbour-subtraction
phase at the ePSF-perturbation stage, using the AIC to find the
location of the close-by neighbours to subtract, before tabulating the
normalised ePSF residuals \citepalias[see][]{Lib16}. For the regions
not covered by the AIC we just collected the ePSF residuals
subtracting only the most obvious neighbour stars clearly visible in
the reconstructed \ktwo exposures.

It is worth mentioning that, as stated in \citetalias{Lib16}, our
ePSFs are still not perfect and a non-negligible room for improvements
is expected when the pixel-response-function calibration data will be
publicly available.

Finally, we measured positions and fluxes of all sources in each \ktwo
reconstructed full-frame exposure with a least-square fit of the
ePSF. We then made a common reference frame system (master frame) for
each channel by cross-identifying all bright, unsaturated stars from
each \ktwo image. Position and flux of a given star in the master
frame were iteratively computed as the clipped average of the
positions and fluxes of that star as measured in each \ktwo exposure
and transformed with six-parameter linear transformations and
zero-point registration into the master-frame reference system.

%%%%%%%%
\section{\ktwo photometry}
%%%%%%%%

We extracted the LCs for most of the objects imaged on module-14 TPFs
during \ktwo/C5. Hereafter, we discuss the key ingredients of our
method.

%%%%%%%%%%%
\subsection{Modified AIC}
%%%%%%%%%%%
\label{maicsub}

As shown in Fig.~\ref{fig1}, the AIC does not completely cover the
entire module-14 FoV, leaving part of channels 45, 46 and 48 partially
unexplored. For this reason, we chose to add the most\footnote{Some
  stars are imaged close to the corresponding TPF boundaries,
  preventing us to perform the PSF fit and measure their positions and
  fluxes.} of the missing stars to the AIC using the \ktwo images
themselves, extracting position and flux of these objects as described
in Sect.~\ref{k2data}. Different factors (e.g., photometric
zero-points and geometric distortion) may vary across such a large
FoV, therefore we performed the procedure described below
independently for each channel.

First, we transformed the position of each missing star from a given
\ktwo exposure (reconstructed using all TPFs with cadence number
108564) into that of the AIC by using six-parameter linear
transformations and added them to the input catalogue. The positions
of the added stars are less precise than those of the stars contained
in the original AIC. Furthermore, for these added stars we do not have
any control about the light-dilution effects. However, in this way we
were able to add, on average, about 130 stars to each channel input
list.

Then, we registered the AIC $N$-filter magnitudes into the
\kmag-magnitude system of the previously-built \ktwo master frame. In
first approximation, the Asiago Schmidt $N$ filter is rather similar
to the \kep total transmission curve, and in \citetalias{Lib16} we
used a simple zero-point to transform the AIC white-light magnitudes
into \kmag magnitudes. However, M\,44 stars are spread over a wide
range of colours in the colour-magnitude diagram (CMD), i.e., $\Delta
(B-I) \sim 6$ magnitudes from the upper to the lower MS, and we found
that such zero-point was not the same for all colours. Therefore, we
performed a photometric calibration by using the ($B-I$) colour to
transform the $N$-filter measurements into \kmag magnitudes. We
applied a least-square fit to find the coefficients of the polynomial
to use for such photometric calibration. If either $B$ or $I$
magnitudes were not available for a given star, we adopted the average
zero-point between $N$-filter and \kmag magnitudes. For the added
stars, which magnitudes are already in the \kmag system, we used a
simple zero-point between the \ktwo selected exposure and the \ktwo
master frame to adjust the magnitudes.

At the end of our integration process, we have four modified AICs
(mAICs), one for each channel, that differ each other for the number
of added stars and for the slightly different calibration
equation. Such mAICs were finally used as input lists during the
LC-extraction phase.

%%%%%%%%%%%
\subsection{Light-curve extraction and systematic correction}
%%%%%%%%%%%
\label{LCextr}

For each channel, we extracted the LCs for all objects in the
corresponding mAIC as described in \citet{Nar15,Nar16} and
\citetalias{Lib16}. Briefly, for each target star in our input list we
used six-parameter, global\footnote{Differently from
  \citetalias{Lib16}, we did not use a local approach because of the
  lacking of close-by stars due to the sparse TPF coverage on the
  channels.} linear transformations to convert its mAIC position into
that of each individual \ktwo exposure. Only bright, well-measured
unsaturated stars were used to compute the coefficients of these
transformations. We then measured its flux both in the original and in
the neighbour-subtracted images\footnote{Note that in \ktwo/C5 the sky
  background was already subtracted from the images. As double-check,
  for each channel we computed the average sky-background level and
  subtracted it from \ktwo exposures. As expected, the sky-background
  value was around zero.}. In the latter case, we subtracted from the
image all close-by stars which light contamination would affect the LC
of our target. M\,44 field is rather sparse, however, in some cases,
there are close-by stars for which the light-dilution effects can be
important during the LC analysis. For each star we performed 1-, 1.5-,
2- and 2.5-pixel aperture and PSF-fitting photometry. Hereafter, we
will consider only the neighbour-subtracted LCs.

The LCs were corrected for the different systematic effects that
usually harm \ktwo data. At variance with \ktwo Campaign 0, the
spacecraft drift was smaller. By simply applying the
position-dependent correction of \citetalias{Lib16}, we found that the
result was not very good, in particular since a few day before the
mid-Campaign Argabrightening
event\footnote{\href{http://keplerscience.arc.nasa.gov/k2-data-release-notes.html}{http://keplerscience.arc.nasa.gov/k2-data-release-notes.html}}
when the stars on the CCDs changed drift pattern because of the change
of the relative positioning of the spacecraft with respect to the
Sun. Furthermore, several stars showed long-term effects not
ascribable to intrinsic variability. Therefore, we improved our LC
detrend with respect to our first work and added a new, preliminary
correction. We refer to our companion paper on the same \ktwo/C5 data
(Nardiello et al., MNRAS submitted) focused on the OC M\,67 for a
detailed description of this systematic-correction stage. In a
nutshell, the correction can be summarised as follows.

\begin{figure*}
  \centering
  \includegraphics[width=\textwidth, keepaspectratio]{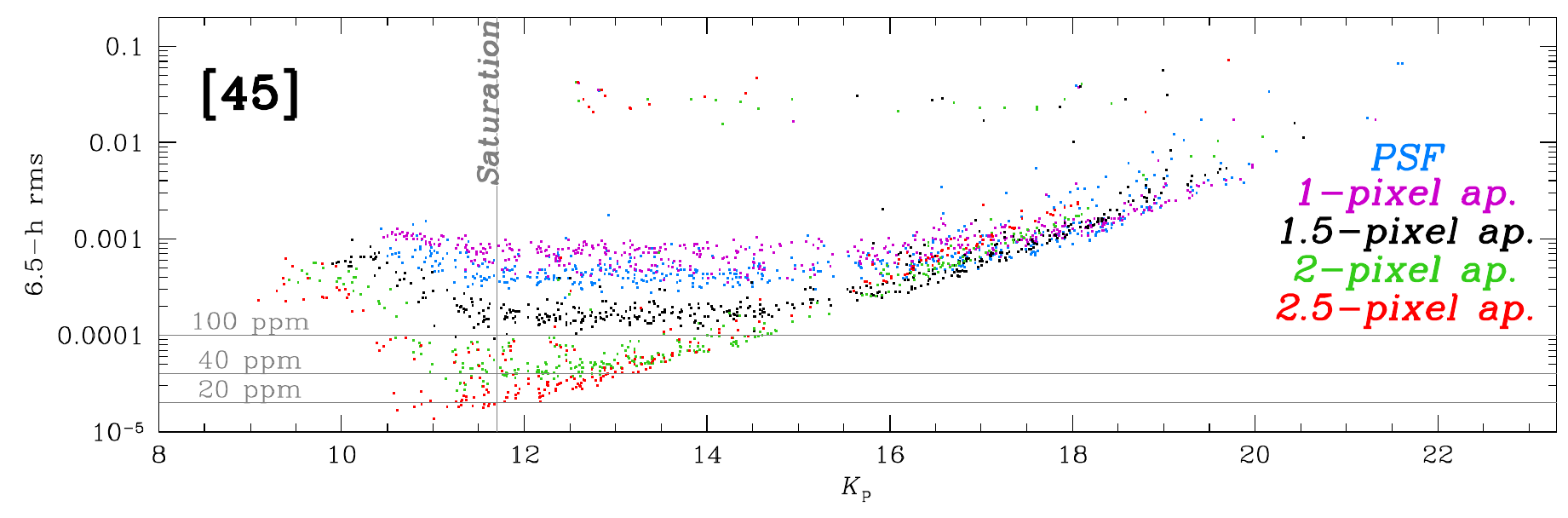}
  \vskip -5.5pt
  \includegraphics[width=\textwidth, keepaspectratio]{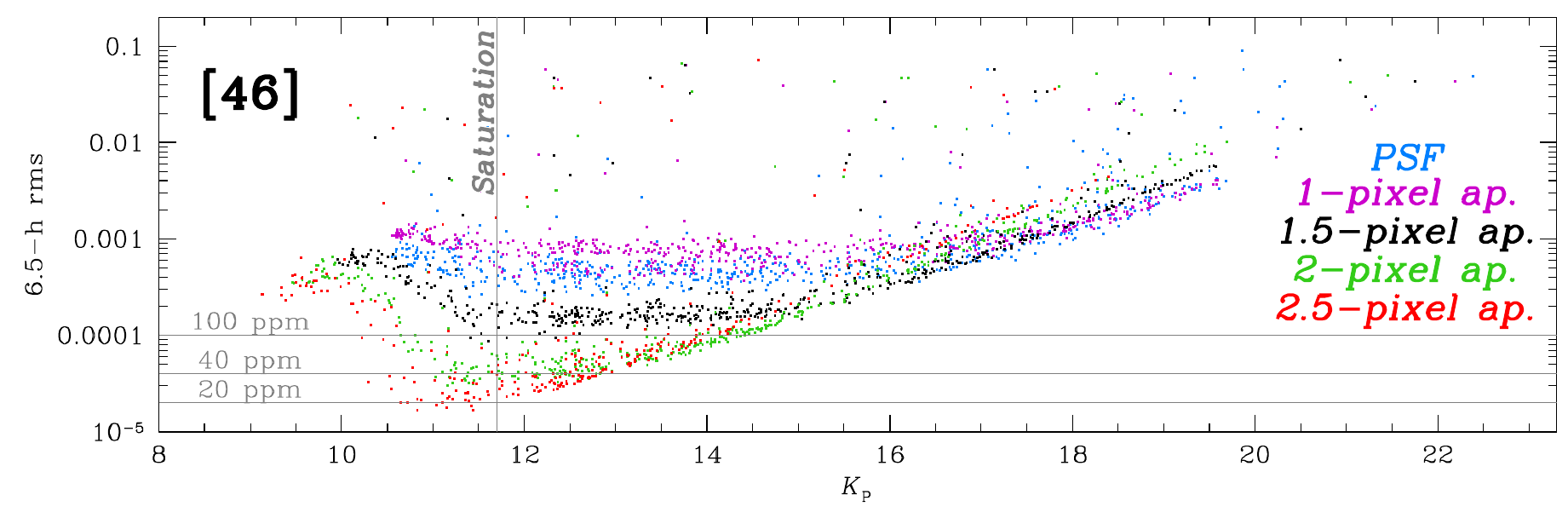}
  \vskip -5.5pt
  \includegraphics[width=\textwidth, keepaspectratio]{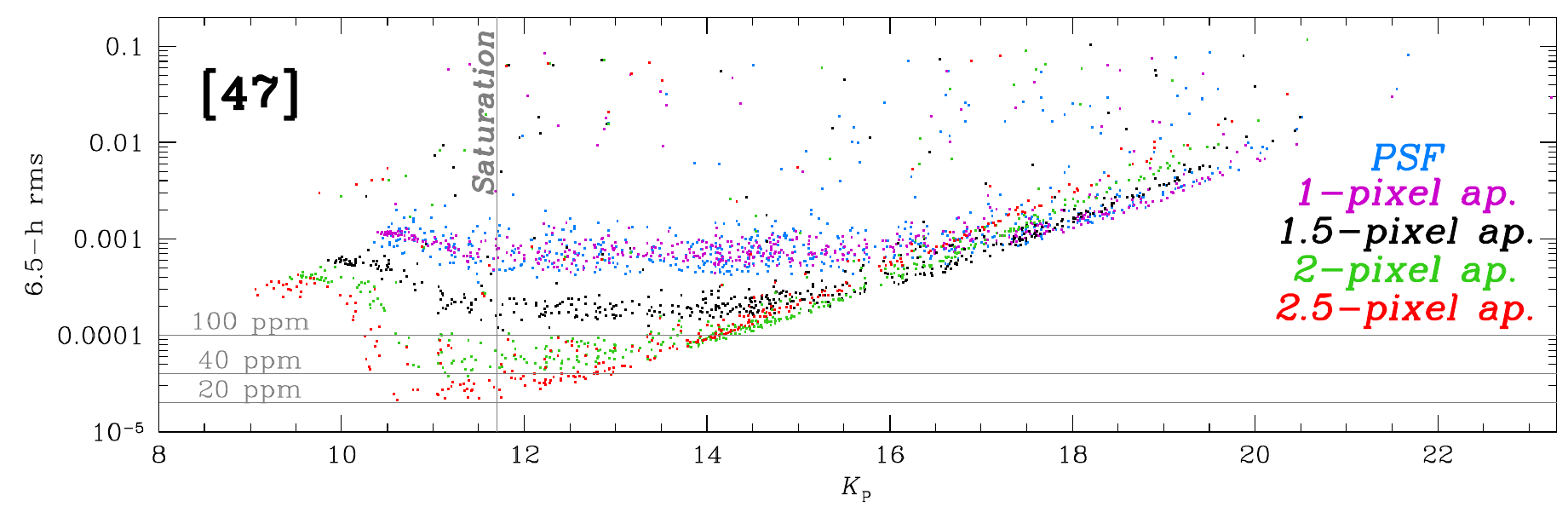}
  \vskip -5.5pt
  \includegraphics[width=\textwidth, keepaspectratio]{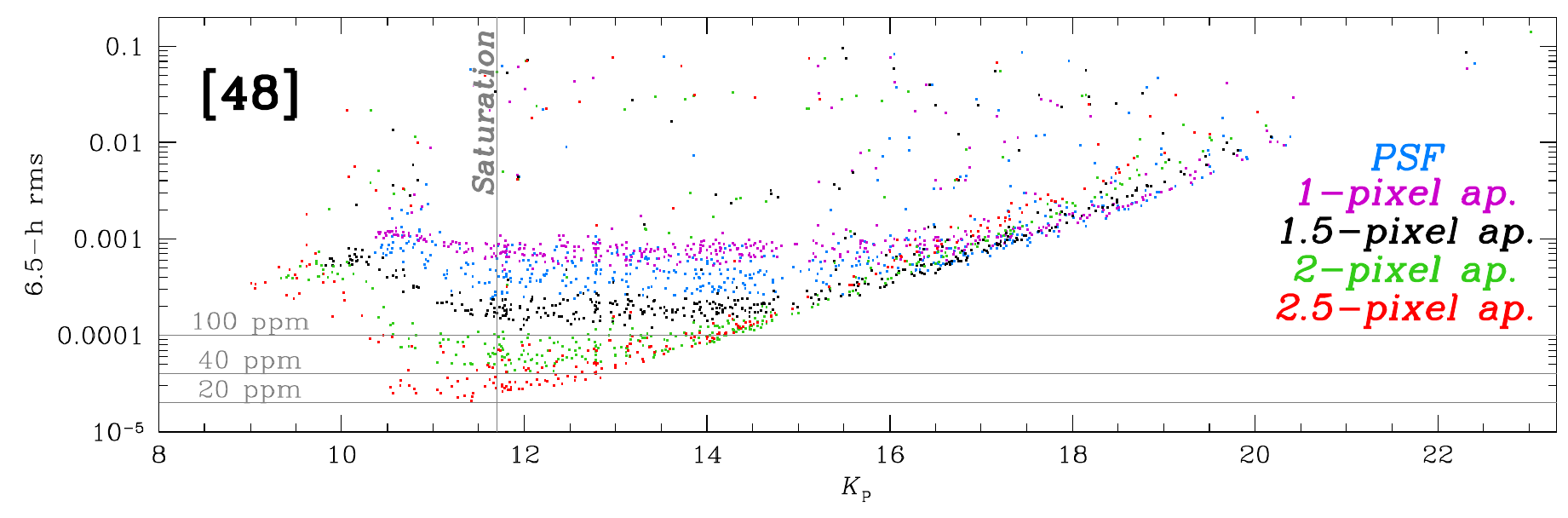}
  \vskip -7.5pt
  \caption{Photometric precision, represented by the 6.5-h rms,
    achieved with 1-pixel-aperture (purple points), 1.5-pixel-aperture
    (black points), 2-pixel-aperture (green points),
    2.5-pixel-aperture (red points) and PSF (azure points) photometry
    on the neighbour-subtracted LCs. We plot the results for each of
    the four channels of \ktwo module 14 separately for clarity. The
    grey, solid horizontal lines are set at 100, 40 and 20 ppm. The
    saturation threshold ($K_{\rm P} \sim 11.7$) is shown with a grey,
    solid vertical line.}
  \label{fig_rms}
\end{figure*}

We first removed the most of the systematic trends that are in common
among the different LCs in a given \ktwo channel. To this task we used
the cotrending basis
vectors\footnote{\href{https://archive.stsci.edu/k2/cbv.html}{https://archive.stsci.edu/k2/cbv.html}}
(CBVs) released with each \ktwo Campaign data set from the third
onward, in a similar way as done by the official \kep pipeline. For
each normalised-flux raw LC we modelled the systematic trends using a
linear combination of the CBVs. The coefficients of such combination
were computed adopting a Levenberg-Marquardt minimisation method
\citep{More80}. For variable stars we noticed that sometimes the
cotrend algorithm tries to include the stellar variability as well in
the CBV linear combination, causing a worsening of the LC. For this
reason, for each star we checked if the LC scatter \citepalias[defined
as the point-to-point, or p2p, rms of][]{Lib16} improved after this
cotrend stage. If not, we used as coefficients of the CBV combination
the average coefficients computed for all stars across the
channel. This way we found an improvement of the LC scatter, even if
sometimes it left some long-term systematics.

The cotrend correction also partially compensated for the
drift-induced trends. However, the correction was based on the common
behaviour of the stars on the CCD, therefore, in order to fine tune
it, we applied to each LC our iterative, position-based detrend as
done in \citetalias{Lib16}. Briefly, we first normalised the raw LC by
its median flux and created a LC model. At odds of \citetalias{Lib16},
the LC model was not obtained with a running-median filter, but with a
linear interpolation. We segmented the LC in different bins and, in
each bin, we computed the 3.5$\sigma$-clipped average flux of the
points. The boundaries of each bin were defined by two consecutive
thruster-jet firings, identified thanks to the ``jumps'' in the
$x$/$y$ raw positions during time. The LC model was generated by
linearly interpolating the LC among these bin average values.
Finally, we removed the correlation between ($x$,$y$) raw positions
and model-subtracted-LC fluxes with a look-up table of correction
applied with a simple bi-linear interpolation. By working with the
model-subtracted LC, we avoided to wrongly correct also the intrinsic
variability of the star.

This correction, in particular the cotrend part, is still in a
preliminary phase. Indeed, we used all 16 CBVs to perform the
correction. In most cases, the correction works very well. However, a
few stars still show residual long-term systematic effects that could
hamper, even if only partially, a variability study. The best solution
should be to check all the possible combinations of CBVs and find
which combination leads to the best photometric precision and
preserves the intrinsic stellar signal. Since such long terms do not
affect the search for eclipsing or transiting objects on these
processed LCs, we postpone the refinement of this cotrend correction
to future works of the series.

%%%%%%%%%%%
\subsection{Photometric precision}
%%%%%%%%%%%
\label{photprec}

In Fig.~\ref{fig_rms} we show the 6.5-h rms \citepalias[defined as
in][]{Lib16} for each of the four analysed channels. Thanks to
observations achieved with a lower spacecraft jitter, the
pixel-to-pixel variations are less effective and the photometric
precision was slightly better than in \citetalias{Lib16}, with a best
value of $\sim$13 parts-per-million (ppm). The \kmag instrumental
magnitudes were registered onto the \kmag system with zero-points (one
for each channel and LC-extraction photometric method) obtained by
comparing our LC-based \kmag instrumental magnitudes with the
EPIC\footnote{\href{https://archive.stsci.edu/k2/epic/search.php}{https://archive.stsci.edu/k2/epic/search.php}}
(Ecliptic Plane Input Catalog) `\textit{gri}'-based \kmag magnitudes
\citepalias{Lib16}.

\begin{figure*}
  \centering
  \includegraphics[width=\textwidth, keepaspectratio]{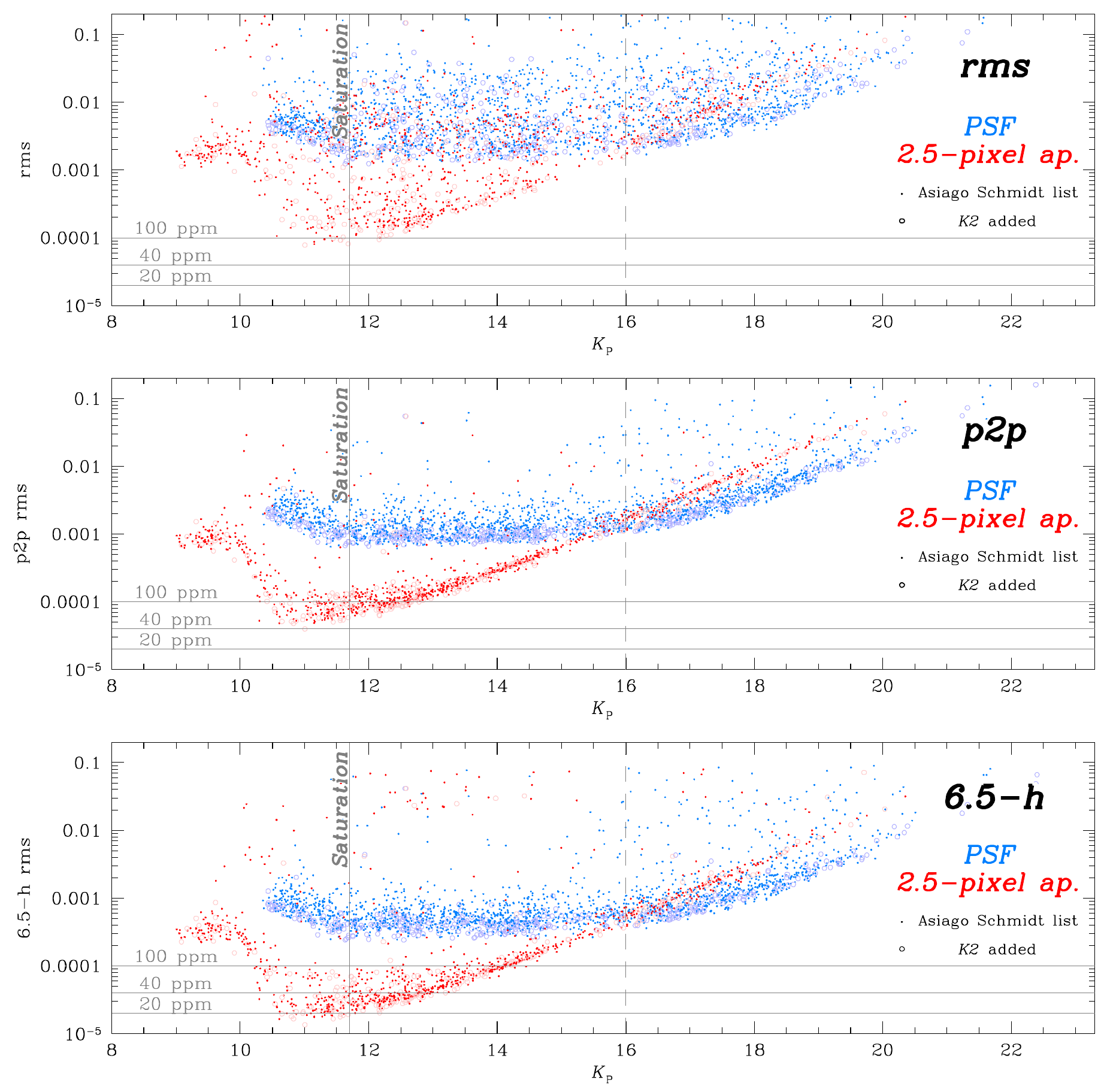}
  \caption{Photometric rms (top panel), p2p rms (middle panel) and
    6.5-h rms (bottom panel) for the 2.5-pixel-aperture- (red points)
    and the PSF-based (azure points) neighbour-subtracted LCs of the
    four module-14 channels together. Stars contained in the original
    AIC are shown with dots, while stars added from \ktwo observations
    are plot with open circles and a lighter colour. The saturation
    threshold is set at $K_{\rm P} \sim 11.7$ (grey, solid vertical
    line). The 100-, 40-, and 20-ppm levels are shown with grey, solid
    horizontal lines. The grey, dashed vertical line at $K_{\rm P}
    \sim 16$ shows where one of the two methods begins to perform
    better than the other.}
  \label{fig_rms_all}
\end{figure*}

As it is clear from Fig.~\ref{fig_rms}, the PSF-based photometry (as
well as the 1-pixel aperture photometry) is more suitable for faint
stars with $K_{\rm P} \gtrsim 17$. This threshold is set $\sim 1.5$
\kmag magnitudes lower than in \citetalias{Lib16}. However, by only
focusing on the 2.5-pixel-aperture (the largest aperture adopted in
this work) and the PSF photometry, the threshold at which one method
overcomes the other is at about $K_{\rm P} \sim 16$, similar to that
found in our first work. In Fig.~\ref{fig_rms_all} we show the simple
rms, the p2p rms and the 6.5-h rms for the 2.5-pixel-aperture and the
PSF photometry in which we collected all module-14 LCs.

In Fig.~\ref{fig_rms_all} we also marked with different symbols stars
included in the original AIC and those that were added to cover the
remaining TPFs outside the AIC FoV. No clear dichotomy arises from the
plot, meaning that our photometric calibrations while building the
mAICs, as well as the registration onto the \kmag system, are good.

%%%%%%%%
\section{Variable-star search}
%%%%%%%%

To find variable stars (e.g., spot-modulated and pulsating stars,
eclipsing binaries, transiting objects), we started by selecting for
each star the LC among the five obtained with the different
photometric methods that shows, on average, the best 6.5-hour rms in
the corresponding magnitude interval. Thruster-jet-related events were
purged from the LC as in \citetalias{Lib16}, while outliers were
removed by performing an asymmetric $\sigma$ clipping\footnote{In
  order to not remove any transit or eclipse event from the LCs we
  proceeded as follows. We first divided the LC in 0.2-d bins and, in
  each bin, we computed the median and the $\sigma$ (defined as the
  68.27$^{\rm th}$ percentile of the distribution around the median)
  values. We then excluded from the subsequent LC analysis all points
  which values were at least $3.5 \sigma$ brighter or $15 \sigma$
  fainter than the median in the corresponding bin.}.

We searched for variable stars using \textsc{VARTOOLS} v1.33 of
\citet{Hart16}. The periodograms were obtained with three different
methods: Generalized Lomb-Scargle \citep[GLS,][]{Press92,Zech09},
Analysis of Variance \citep[AoV,][]{SC89} and Box-fitting Least-Square
\citep[BLS,][]{KZM02}. To detect variable-star candidates, we first
made an histogram of the periods of all the analysed LCs and removed
the spikes that are associated with spurious signals such as
thruster-jet firing or other systematic effects. For GLS and AoV, we
then plotted the SNR as a function of the period and selected by hand
stars that show an high SNR. For BLS we used the signal-to-pink noise
\citep{Pont06} instead of the SNR. A complete description of the
method, supplied with figures, is available in \citet{Nar15} and
\citetalias{Lib16}.

Among the 2199 field and cluster stars for which we extracted a
reliable LC, 1654 objects present a variability signature. As in
\citetalias{Lib16}, we classified them (by eye comparing the LC of
each candidate with those of the close-by neighbours) in three
distinct groups: stars that have a high-probability to be true
variable sources, eclipsing binaries and candidate exoplanets (1494
stars), probable blends (33 stars) and objects that were difficult to
judge just looking at the LC (127 stars). In the latter group there
are true variables, blends and stars for which long-term or residual
systematic effects could be confused for variability (and vice versa).

%%%%%%%%%%%
\subsection{Cross-match with the literature}
%%%%%%%%%%%
\label{matchlit}

To estimate the completeness of our variable catalogue, we matched our
mAICs with several already-published catalogues focused on M\,44. The
works considered in our analysis are the following: \citet{Agu11},
\citet{Bou01}, \citet{Bre12}, \citet{Cas12}, \citet{Del11},
\citet{Dou14}, \citet{Dra14}, \citet{Kov14}, \citet{Li07},
\citet{Liu07}, \citet{Mer09}, \citet{Pep08}, \citet[GCVS]{GCVS},
\citet{Sch11}, and the ``Variable Star Index'' (VSX) catalogue.
    
Of the 1621 (1494 candidate and 127 ``difficult-interpretation'')
variables we have found, 550 objects were contained in other
catalogues. Additional 72 already-known variables were imaged on a TPF
during \ktwo/C5 but were not included in our catalogue, therefore we
visually inspected again these remaining objects and chose whether to
add them or not to our list. In total we added 26 of these missing
stars. The remaining 46 already-known variables were not included in
our catalogue for different reasons. The missing stars (i) are heavily
saturated in these long-cadence images, (ii) are too close to the TPF
boundaries or to a very saturated star, (iii) are not included in our
mAICs, and/or (iv) do not show any variability in the LC (some objects
are listed in catalogues based on spectroscopic/RV observations,
therefore we may not be able to detect any variability/binary
signature in their LCs).

In total we found 1071 (954 candidate and 117
``difficult-interpretation'') new variables in this M\,44 field. We
emphasise that, as discussed above and in Sect.~\ref{LCextr}, some
long-term systematics left after our detrending may be wrongly
interpreted as long-term variability, and hence the new-candidate list
could be shorter. Therefore, 1071 should be considered as an
indicative value.

\begin{figure*}
  \centering
  \includegraphics[height=10cm]{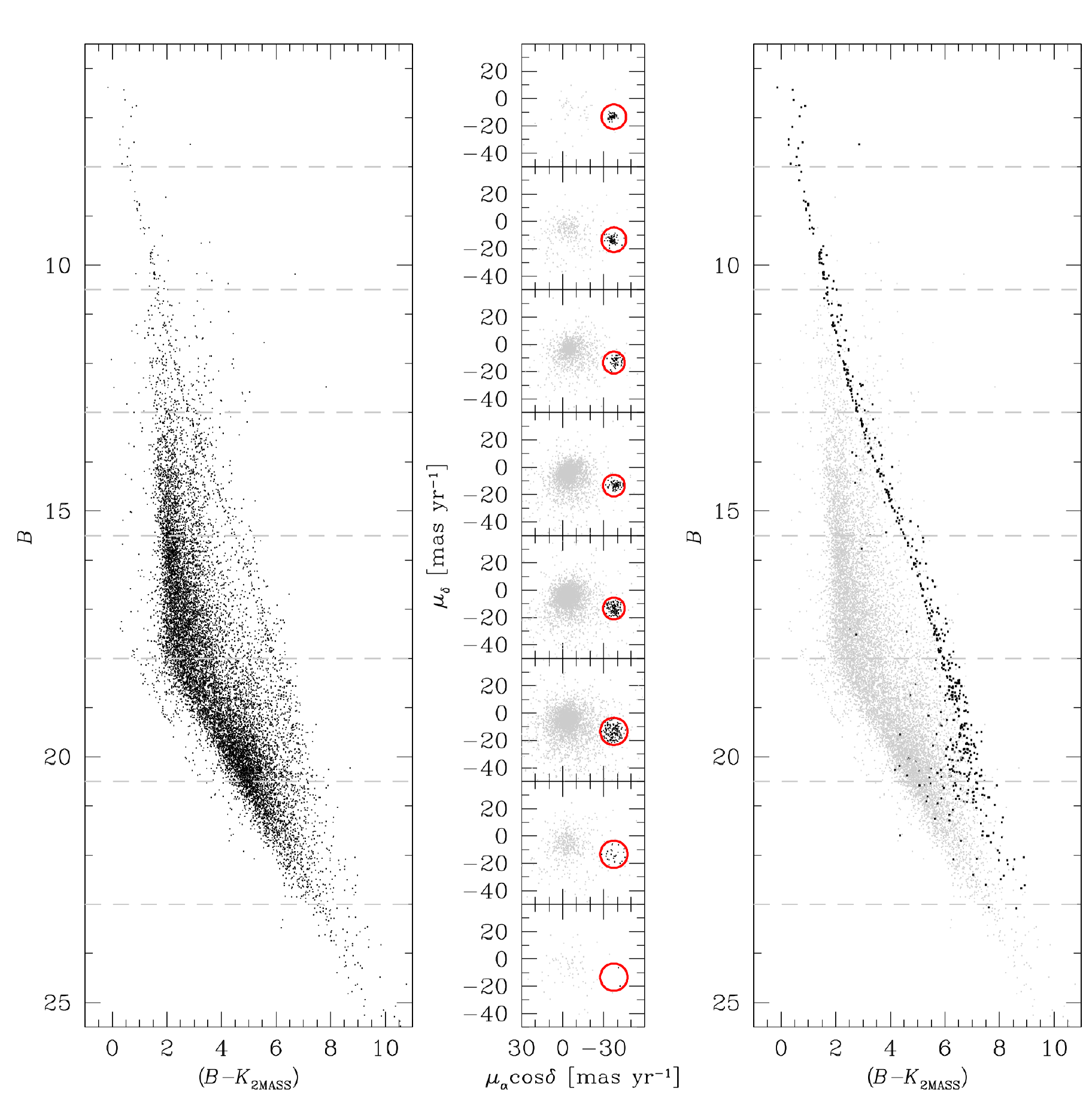}
  \includegraphics[height=10cm]{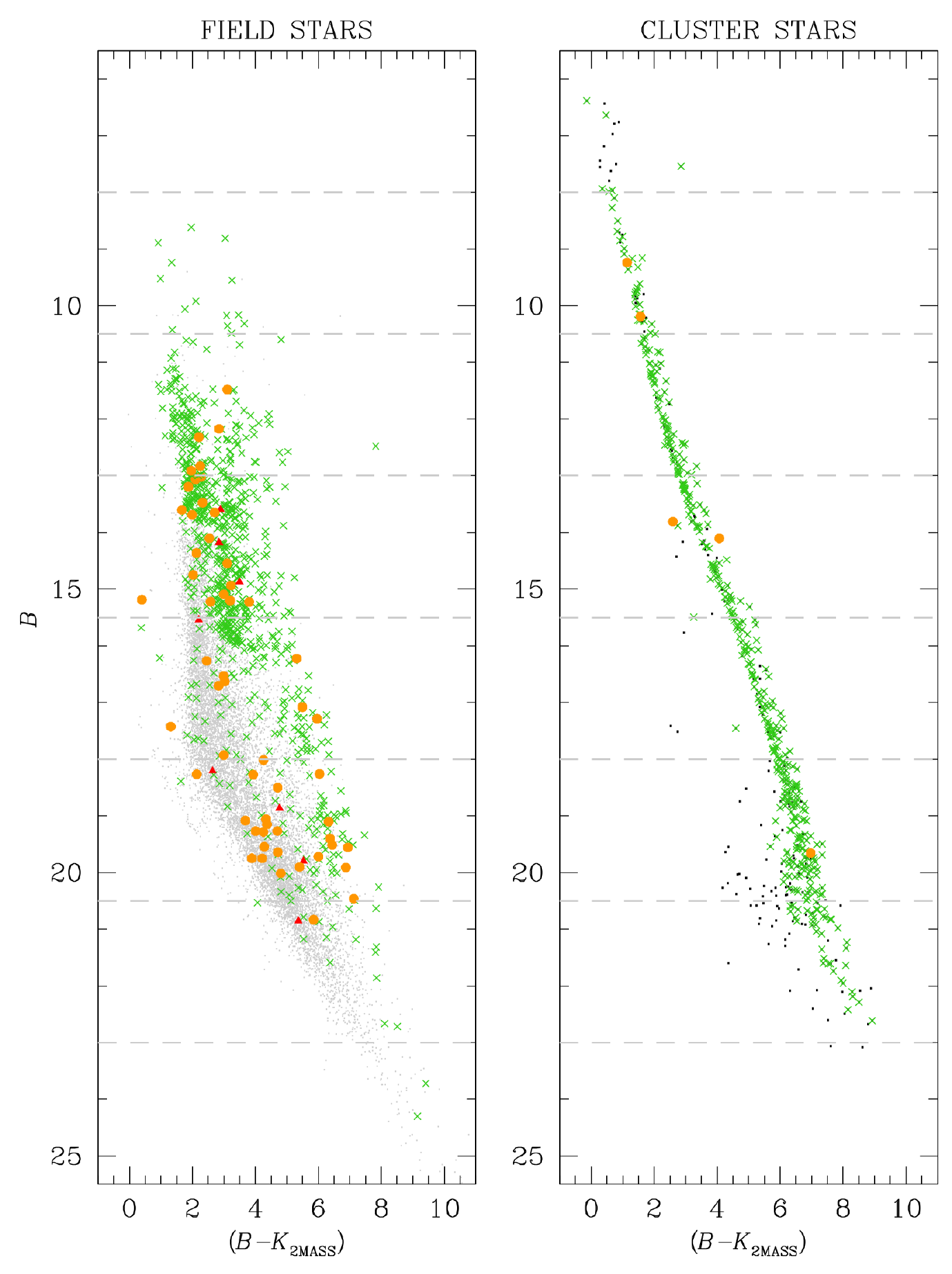}
  \caption{From left to right. First column: $B$ versus ($B-K_{\rm
      2MASS}$) CMD of the stars in the mAICs. We split the CMD in
    eight 2.5-magnitude bins (defined by the horizontal, grey dashed
    lines) to better select the cluster members using PPMXL proper
    motions. Second column: vector-point diagrams for each magnitude
    bin. The M\,44 members distribution is centred around
    ($-37.47,-13.39$) mas yr$^{-1}$. Field (grey dots) and cluster
    (black dots) stars were selected accordingly to their location
    (outside or inside, respectively) with respect to the red circles
    centred on the M\,44 distribution. The radius of these circles
    ranges from a minimum of 8 mas yr$^{-1}$ to a maximum of 10 mas
    yr$^{-1}$. Third column: same CMD as on the first column but with
    stars plot colour-coded as in the previous vector-point
    diagrams. Thanks to our proper-motion-based selection, we are able
    to clearly separate cluster from field stars. Fourth column: CMD
    with only field stars in which we highlighted the detected
    candidate variables (green crosses), the
    ``difficult-interpretation'' objects (orange dots) and the blends
    (red triangles). Fifth column: as in the fourth column but for
    M\,44 members.}
  \label{fig_cmd}
\end{figure*}

%%%%%%%%%%%
\subsection{CMDs and vector-point diagrams}
%%%%%%%%%%%

After the cross-match with the literature, we have a catalogue with
1680 stars: 1520 candidate variables, 127 ``difficult-interpretation''
objects and 33 blends. In Fig.~\ref{fig_cmd} we show the $B$
vs. ($B-K_{\rm 2MASS}$) colour-magnitude diagrams (CMDs) and the
vector-point diagrams of the stars observed in our mAICs (for which we
have a PPMXL-proper-motion, a $B$- and a $K_{\rm 2MASS}$-magnitude
value). The proper-motion selections were made similarly as in
\citet{Lib15}. First, we divided the CMD in eight 2.5-magnitude bins
and, for each of such bin, we drew a circle in the corresponding
vector-point diagram to select only stars with a cluster-like
motion. The adopted radius was chosen as a compromise between
excluding cluster members with poorly-measured proper motions and
including field stars lying in the cluster-bulk locus. Among the
variables in our catalogue with a proper-motion measurement, we found
that $\sim$32\% of them have a high-probability to be cluster members,
while the remaining $\sim$68\% of the stars belong to the field in the
direction of M\,44.

%%%%%%%%%%%
\subsection{Peculiar objects}
%%%%%%%%%%%

In our analysis we have found two peculiar objects that are worth to
be mentioned. In the following subsections we briefly describe them.

%%%%%%%%%%%%%%
\subsubsection{LC \# 24092 - Channel 45}
%%%%%%%%%%%%%%

\begin{figure*}
  \centering
  \includegraphics[width=\textwidth]{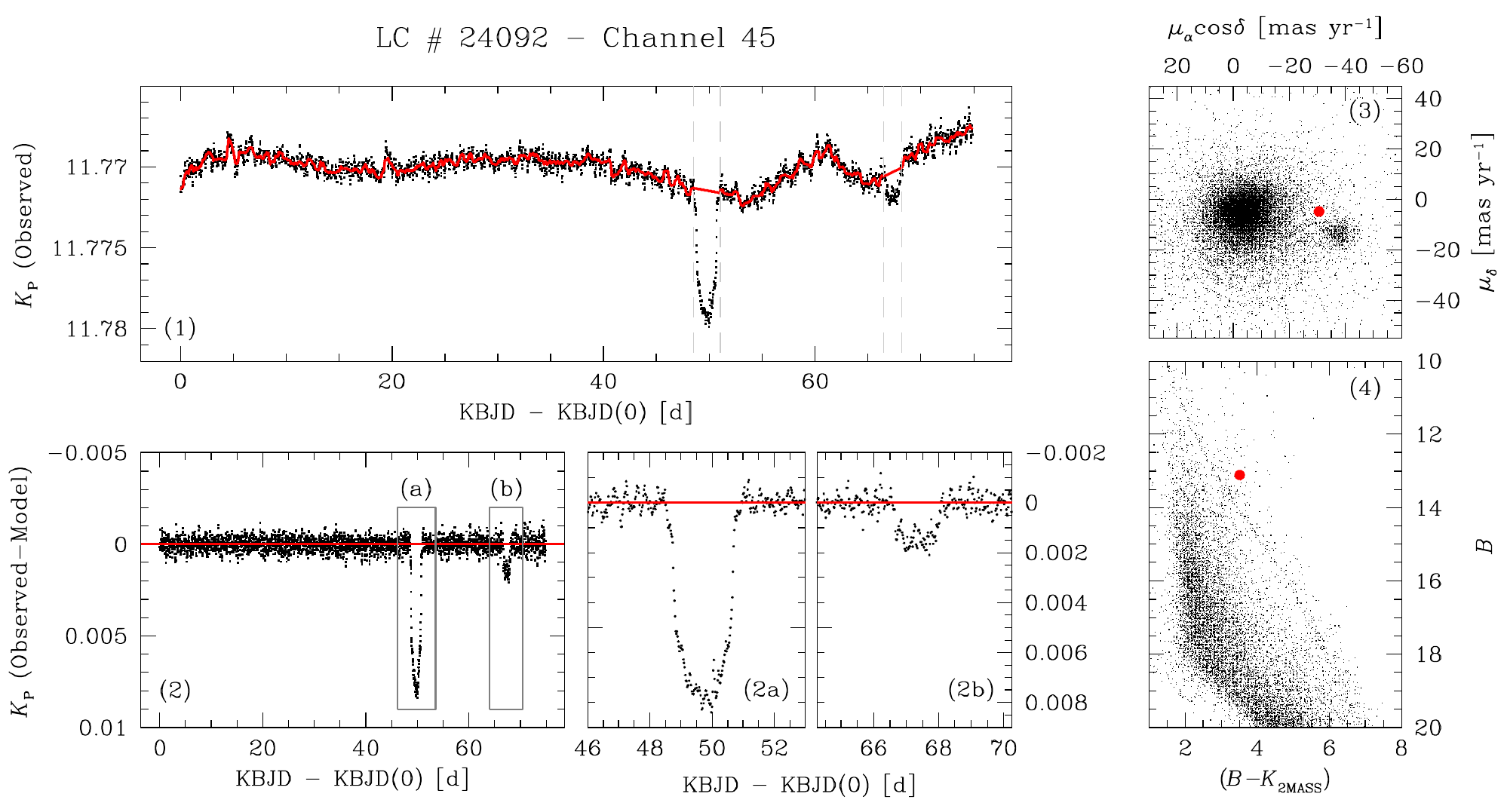}
  \caption{Overview of star \# 24092 in channel 45
    (EPIC~211892898). In panel (1) we show the original, detrended
    2.5-pixel-aperture LC. The red line represents the LC model
    obtained using a running-median filter with window of 6 hours. The
    two eclipses/transits (between the grey, dashed, vertical lines)
    were excluded during the LC modelling. In panel (2) we show the
    difference between observed and model LCs. Panels (2a) and (2b)
    highlight the two supposed eclipses/transits. Finally, on the
    right panels we show the vector-point diagram (panel 3) and the
    $B$ vs. ($B-K_{\rm 2MASS}$) CMD (panel 4). The red dot marks the
    location of star \# 24092 - 45 in each panel.}
  \label{fig:24092}
\end{figure*}

Star \# 24092 - 45\footnote{($\alpha$,$\delta$)$_{\rm
    J2000.0}$$\sim$($130^{\circ}\!\!.89305,+18^{\circ}\!\!.622461$)}
(EPIC~211892898) is an eclipsing or transiting field object with a
period greater than 50 d. By adopting the stellar parameters given by
the \ktwo \textsc{EXOFOP}
website\footnote{\href{https://exofop.ipac.caltech.edu/k2/index.php}{https://exofop.ipac.caltech.edu/k2/index.php}},
this object is an eclipsing binary for which we can detected only one
primary and one secondary eclipse, and therefore it may have an
eccentric orbit. The primary-eclipse depth is of $\sim$0.085 \kmag
magnitude, while the secondary eclipse has a depth of $\sim$0.002
\kmag magnitude. As shown in Fig.~\ref{fig:24092} the eclipses last
for about two days, suggesting that the system is almost edge-on, with
the two components that have a large radii difference and/or that are
far from each other. The hypothesis of a grazing eclipsing binary with
a large eccentricity cannot be discarded as well.

If the true mass and radius of the star are different than
\textsc{EXOFOP} values, a possible interpretation is that we are
looking at a planetary system. A RV follow-up is required in order to
shed light on the true nature of this object.

%%%%%%%%%%%%%%
\subsubsection{LC \# 24175 - Channel 46}
%%%%%%%%%%%%%%

\begin{figure*}
  \centering
  \includegraphics[width=\textwidth]{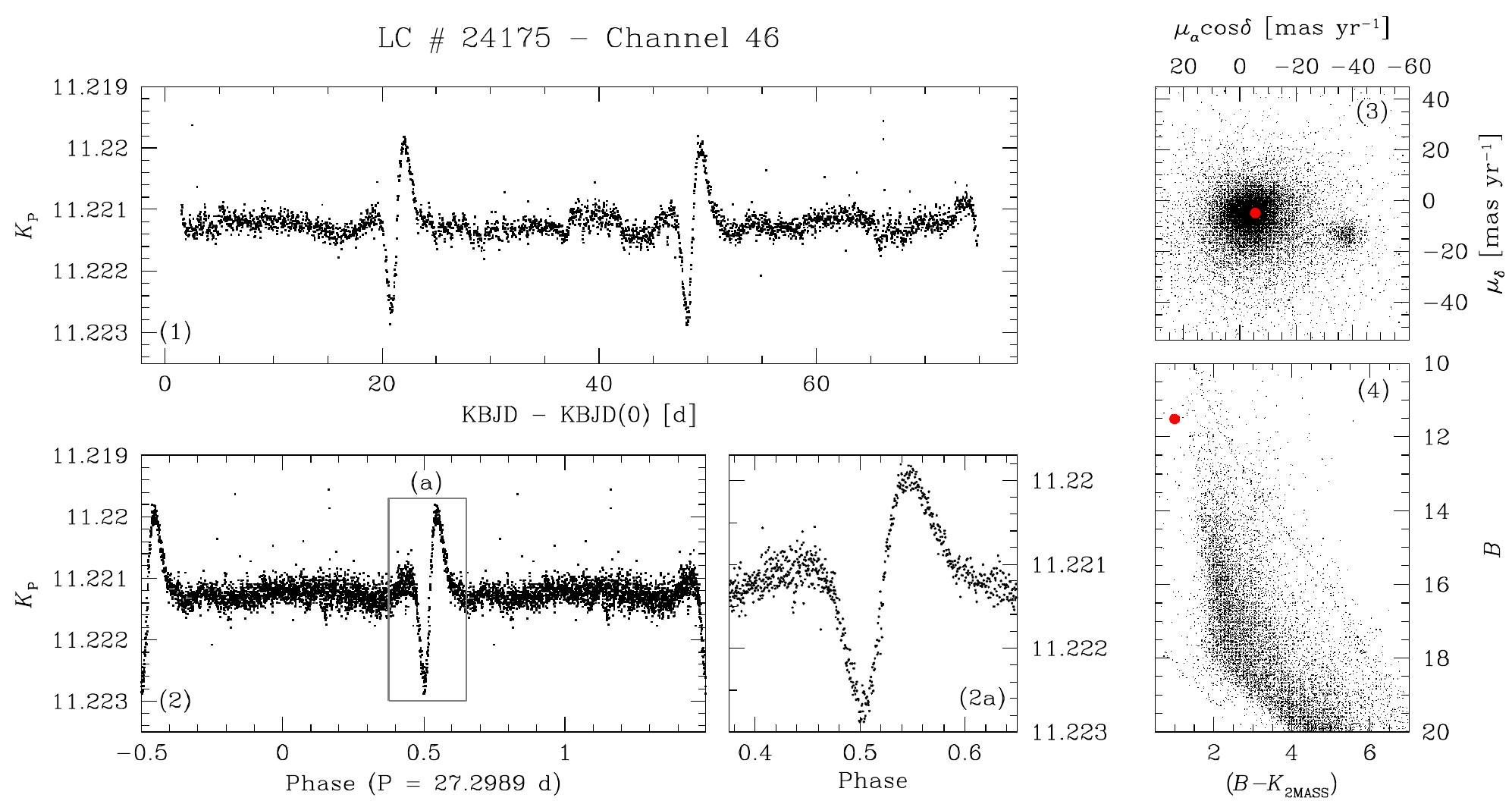}
  \caption{Overview for the star \# 24175 in channel 46
    (EPIC~211896553). On the left panels we plot the original,
    detrended 2.5-pixel-aperture LC (panel 1), the phased LC with a
    period of $\sim$27.3 d (panel 2) and a zoom-in around the
    heartbeat (panel 2a). In panels (3) and (4) we show the
    vector-point diagram and the $B$ vs. ($B-K_{\rm 2MASS}$) CMD,
    respectively. Similarly to Fig.~\ref{fig:24092}, in the right
    panels the red dot marks the location of star \# 24175 - 46.}
  \label{fig:24175}
\end{figure*}

Star \# 24175 - 46\footnote{($\alpha$,$\delta$)$_{\rm
    J2000.0}$$\sim$($129^{\circ}\!\!.07402,+18^{\circ}\!\!.676282$)}
(EPIC~211896553) is a potential eccentric binary known as heartbeat
binary \citep[e.g.,][]{Thom12} not member of M\,44. The heartbeat
shape in the LCs of these objects is due to tidal distortion of the
star after a fly-by at the periastron that changes its
brightness. These rare systems (173 heartbeat stars currently known)
are characterised by large eccentricities and periods between a
fraction of day and $\sim$450 d \citep[see][]{Kirk16}. Our detected
binary has a period of $\sim$27.3 d (Fig.~\ref{fig:24175}). RV
measurements are required to constrain the orbital parameters and
model the system.

%%%%%%%%
\section{Exoplanet search}
%%%%%%%%
\label{exoplanets}

We searched for candidate exoplanets in M\,44 field. To this purpose,
we applied a specific procedure that can be summarised as follows.

For each LC, we initially modelled all residual long-term systematic
effects and intrinsic variability of the star using a 3$^{\rm
  th}$-order spline with 150 break points and subtracted such model
from the LC. We also removed the most of the outliers with an
iterative $\sigma$ clipping. Hereafter, we will label the
model-subtracted LCs as ``flattened'' LCs.

For each flattened LC, we extracted the periodogram using
\textsc{VARTOOLS} BLS task (searching for periods between 0.5 and 75
d) and normalised it as done by \citet{Vander16} to decrease the
number of false detections in the long-period regime of the
spectrum. We then iteratively selected the five peaks with the highest
SNR, every time excluding from the subsequent selection all harmonics
with $P = N \cdot P_{{\rm sel}}$ and $P = 1/N \cdot P_{{\rm sel}}$,
$N$ integer. We also avoided to include spurious frequencies (e.g.,
those related to the spacecraft jitter) in our selection.

For each selected period $P_{\textrm{sel},i=1,...,5}$, we searched for
a significant flux drop in the flattened LC. First, we run
\textsc{VARTOOLS} BLS, this time fixing the period to refine the
central time and the duration of the possible transit. We then phased
the flattened LC, computed the median magnitude at the centre of the
transit (we used only points within a half of the transit duration
centred at the mid-transit time) and checked whether this value was at
least 1 $\sigma$ (defined as the 68.27$^{\rm th}$ percentile of the
distribution around the median) below the out-of-transit level or
not. If it was not, we discarded the candidate. Next, we verified that
the flux drop found was not due to some outliers by comparing the
median magnitude at the centre of the transit with and without
considering any point within 1 $\sigma$ from the out-of-transit
level. If the two values were comparable, the flux drop was
real. Finally, we investigated if the examined period was correct or
not, by checking for similar flux drops at different phases in the
folded flattened LC. For this exoplanet search we chose to also rely
on diagnostics other than those provided by BLS because, by setting a
selection threshold based on BLS outputs alone, we could exclude good
candidates with very shallow transits and noisy LCs, as well as
include too many false detections (e.g., eclipsing binaries or
RR-Lyrae stars) with high SNR in BLS.

\begin{figure*}
  \centering
  \includegraphics[width=\textwidth, keepaspectratio]{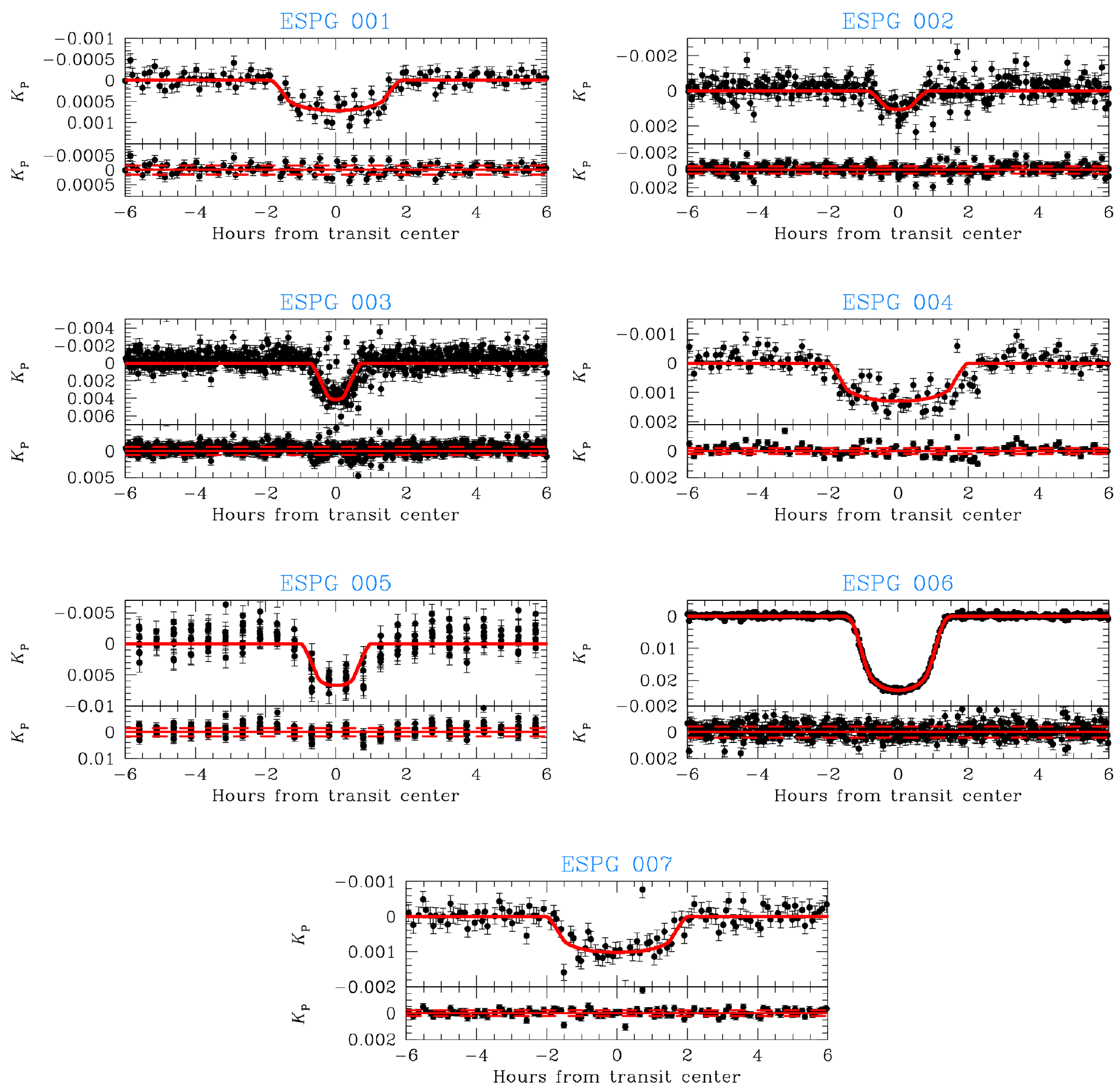}
  \caption{Overview of the seven exoplanet candidates discovered in
    this work. For each candidate, on top we plot the phased flattened
    LC (black dots) with the corresponding model (red solid line)
    obtained as described in Sect.~\ref{vandm}. The size of the error
    bars are computed as the 68.27$^{\rm th}$ percentile of the
    distribution of the out-of-transit points around the median. On
    bottom we show the difference between data and model. The
    horizontal, red solid line is set at 0, while the two horizontal,
    red dashed lines are set at $\pm$68.27$^{\rm th}$ percentile of
    the distribution of the residuals around the median.}
  \label{espg_over}
\end{figure*}

We also double-checked the goodness of the five periods by performing
the same analysis described above using multiples ($N$$=$2,3,4) and
sub-multiples ($N$$=$$1/2$,$1/3$,$1/4$) of
$P_{\textrm{sel},i=1,...,5}$ to take into account that one of the
selected peaks could actually be an harmonic of the true period. In
total, we explored 35 periods per LC. All stars that passed the
aforementioned criteria for one out of the 35 selected periods were
visually inspected in their phased LCs and eventually selected for the
final candidate sample. In total we detected seven transiting
exoplanet candidates. An overview of the candidates is presented in
Fig.~\ref{espg_over}. Single-transit objects were not considered in
our analysis.

%%%%%%%%%%%
\subsection{Vetting and modelling}
%%%%%%%%%%%
\label{vandm}

To verify the goodness of the seven candidates, we first visually
inspected the LCs of the variable stars within 100 \kep pixels from
each candidate to search for possible eclipsing binaries miming the
transit event and found that these candidates are sufficiently
isolated.

We then investigated a possible correlation between transit events and
stellar position. To this task, we cannot use the position of the star
in the raw-image reference frame because of the \kep pointing jitter
that moves the stars on the CCD and harms the true comprehension of
any possible correlation between the transit event and the star
location. Therefore we needed an ``absolute'' reference frame in which
we can safely compare the positions and we chose that of the mAIC. The
position of the star in the mAIC reference-frame system at each epoch
was computed as follows. In each \ktwo/C5 image, we subtracted all
close-by neighbours to the exoplanet candidate as done during the LC
extraction. We then estimated the position of the star by PSF fitting
and transformed it onto the mAIC reference frame by inverting the
six-parameter, global linear transformations adopted during the LC
extraction (see Sect.~\ref{LCextr}). Within the transformation and the
geometric-distortion errors, for our seven candidates no clear
correlations arise (panel f in Fig.~\ref{espg_1_2}, \ref{espg_3_4},
\ref{espg_5_6} and \ref{espg_7}).

After these validations, we fitted a transit model to extract the
transit parameters of these candidates. In order to have a preliminary
estimate, we combined the particle-swarm algorithm
\texttt{Pyswarm}\footnote{Modified version of the public-available
  code at
  \href{https://github.com/tisimst/pyswarm}{https://github.com/tisimst/pyswarm}}
with the \citet{MandA02} model implemented in
\texttt{PyTransit}\footnote{\href{https://github.com/hpparvi/PyTransit}{https://github.com/hpparvi/PyTransit}}
\citep{Parv15}, and used the
\texttt{emcee}\footnote{\href{http://dan.iel.fm/emcee/current/}{http://dan.iel.fm/emcee/current/}
  and
  \href{https://github.com/dfm/emcee}{https://github.com/dfm/emcee}}
algorithm \citep{FM13} to compute the corresponding errors. For each
candidate we adopted the stellar parameters (mass and radius) provided
by \citet{Hub16}, retrieved from the \ktwo \textsc{EXOFOP} website.

We started by purging the most of the outliers from the flattened LCs
in order to avoid to model spurious artefacts. We adopted three
different methods, tailored for each candidate, to obtain the best
purged LC for the subsequent analysis: (i) we subtracted a crude
transit model and performed a 3-, 5- or 10-$\sigma$ clipping in the
observed-minus-model plane; (ii) we selected only transit
neighbourhoods and discarded the off-transit parts of the LC; (iii) we
combined the previous two approaches.

In our transit modelling we made some assumptions. We fixed the
eccentricity ($e$) and the argument of pericentre ($\omega$) to 0 and
90 deg, respectively. For the limb darkening, we chose a quadratic law
and computed the linear and quadratic coefficients with \textsc{JKTLD}
\citep{Sou08} that makes use of the table of \citet{Sing10}. As input
for \textsc{JKTLD}, we adopted $T_{\rm eff}$, $\log g$ and [M/H]
released in \textsc{EXOFOP}, while the microturbolence velocity was
fixed at 2 km s$^{-1}$.

The only values that we chose to characterise were the period ($P$),
the mid-transit time of reference ($T_0$), the inclination ($i$) and
the radii ratio ($R_{\rm P} / R_{\rm S}$). For each parameter, we set
specific limit values within which to search for the best estimate. We
defined $P$ and $T_0$ boundaries around guess values obtained by
running \textsc{VARTOOLS} BLS. The inclination and the radii ratio
were allowed to span a wide range of values, between 70 deg and 110
deg for $i$ and between $10^{-4}$ and $0.5$ for $R_{\rm P} / R_{\rm
  S}$.

Once set the parameter limits, we let the \texttt{Pyswarm} algorithm
span within the boundaries with 180 different parameter configurations
for 10\,000 iterations. We evaluated the goodness of the fit as the
reduced chi-square:

\begin{equation}
  \label{eq:chi2r}
  \chi^{2}_\textrm{r} = \frac{\chi^{2}}{\rm dof} = \sum_{j=1}^{N}{\left(\frac{O_j - M_j}{\sigma_O}\right)} \cdot \frac{1}{\rm dof} \ ,
\end{equation}

\noindent where $j=1,...,N$ with $N$ number of data points in the LC,
$O_j$ and $M_j$ are the observed and the model data-point value,
$\sigma_O$ is the associated error equal to the intrinsic
dispersion\footnote{Defined as the 68.27$^{\rm th}$ percentile of the
  residuals with respect to the median value of the flattened LC.} of
the flattened LC, and ``dof'' means degrees of freedom (the difference
between the number of data points and the number of fitting
parameters). Then, we took the 60 best combinations to initialise the
walkers for the \texttt{emcee} algorithm. For each fitting parameter
we used uniform priors within the same boundaries defined in
\texttt{Pyswarm}. We let the 60 walkers evolve for 40\,000 steps,
maximising the log-likelihood defined as:

\begin{equation}
  \label{eq:lglkhd}
  \ln \mathcal{L} = - \frac{\chi^{2}}{2} \ .
\end{equation}

The best parameters were computed as the median values of the
posterior distributions after conservatively discarding as burn-in
phase the first 10\,000 steps to ensure the convergence of the
chains. The related errors were defined as the 68.27$^{\rm th}$
percentile of the absolute residuals with respect to the median of the
posterior distributions of the fitted parameters.

In Table~\ref{Tab:exo} we list all parameters obtained by our LC
modelling. Again, we emphasise that our model is strongly dependent on
the (fixed) stellar parameter we adopted. A more-reliable estimate of
the exoplanet parameters may be obtained after a RV follow-up (at
least for the brightest targets). As reference, we also report the
photometric transit depth $\delta_{\rm Phot}$. This value was computed
as follows. We normalised the LC by its median flux and phased it
using the period given by our previous analysis. We set the transit
centre at $\rm Phase = 0.5$ and computed $\delta_{\rm Phot}$ as the
median value of the points with $\lvert \rm Phase - 0.5 \rvert <
0.004$. The related error was computed as:

\begin{equation}
  \sigma_{\rm \delta_{Phot}} = \sqrt{\frac{\sigma_{\rm in}^2}{\sqrt{N_{\rm in}-1}} + \frac{\sigma_{\rm out}^2}{\sqrt{N_{\rm out}-1}}}  \ ,
\end{equation}

where $\sigma_{\rm in}$ and $\sigma_{\rm out}$ are the 68.27$^{\rm
  th}$ percentile of the distribution around the median for the points
in ($\lvert \rm Phase - 0.5 \rvert < 0.004$) and out of transit, and
$N_{\rm in}$ and $N_{\rm out}$ are the number of points used in the
calculation.

\begin{landscape}
\begin{table}
  \caption{Exoplanet-candidate parameters.}
  \centering
  \label{Tab:exo}
  \fontsize{7.5pt}{5pt}\selectfont
  \begin{threeparttable}
  \begin{tabular}{cccccrcccccc}
    \hline
    \hline
    & \\
    \textbf{Candidate} & \textbf{EPIC} & \textbf{R.A.} & \textbf{Dec.} & \textbf{$K_{\rm P}$} & \multicolumn{1}{c}{\textbf{Period}} & \textbf{$T_0$} & $i$ & \textbf{$R_{\rm P} / R_{\rm S}$} & \textbf{$\delta_{\rm Phot}$} & \textbf{$R_{\rm S}$} & \textbf{$R_{\rm P}$} \\
    & & [deg] & [deg] & & \multicolumn{1}{c}{[d]} & [KBJD] & [deg] & & [\%] & [$R_\odot$] & [$R_{\rm Jup}$] \\
    & \\
    \hline\hline
    & \\
    ESPG 001 & 211913977 & 130.34349 & $+18.934026$ & 12.646 & $14.675828 \pm 0.000670$ & $2319.686540 \pm 0.001737$ & $89.26 \pm 0.07$ & $0.0233 \pm 0.0004$ & $0.0638 \pm 0.0033$ & 0.725 & 0.165 \\
    & \\
    ESPG 002 & 211897691 & 130.08191 & $+18.693113$ & 14.323 & $\phantom{0}5.749481 \pm 0.000240$ & $2309.495219 \pm 0.001717$ & $86.89 \pm 0.04$ & $0.0351 \pm 0.0013$ & $0.0753 \pm 0.0091$ & 0.765 & 0.261 \\
    & \\    
    ESPG 003 & 211924657 & 130.02655 & $+19.092411$ & 15.048 & $\phantom{0}2.644259 \pm 0.000047$ & $2309.002608 \pm 0.000650$ & $90.00 \pm 0.12$ & $0.0563 \pm 0.0005$ & $0.3274 \pm 0.0167$ & 0.225 & 0.123 \\
    & \\
    ESPG 004 & 211919004 & 129.77680 & $+19.010098$ & 13.135 & $11.722228 \pm 0.000439$ & $2316.084281 \pm 0.001422$ & $90.00 \pm 0.09$ & $0.0308 \pm 0.0002$ & $0.1296 \pm 0.0055$ & 0.799 & 0.240 \\
    & \\
    ESPG 005 & 211916756 & 129.36243 & $+18.976653$ & 16.172 & $10.134231 \pm 0.000347$ & $2317.876813 \pm 0.001248$ & $90.00 \pm 0.06$ & $0.0713 \pm 0.0010$ & $0.5276 \pm 0.0241$ & 0.226 & 0.157 \\
    & \\
    ESPG 006 & 211929937 & 129.17834 & $+19.173816$ & 14.165 & $\phantom{0}3.476633 \pm 0.000006$ & $2309.412293 \pm 0.000074$ & $87.74 \pm 0.01$ & $0.1341 \pm 0.0001$ & $2.0868 \pm 0.0098$ & 0.865 & 1.130 \\
    & \\
    ESPG 007 & 212008766 & 129.28246 & $+20.399322$ & 12.822 & $14.130142 \pm 0.000844$ & $2312.117423 \pm 0.002062$ & $89.44 \pm 0.13$ & $0.0278 \pm 0.0004$ & $0.0942 \pm 0.0040$ & 0.794 & 0.215 \\
    & \\
    \hline
  \end{tabular}       
  \begin{tablenotes}
  \item \textbf{Notes.} \kmag is the median magnitude in the
    LC. Period, $T_0$, $i$ and $R_{\rm P} / R_{\rm S}$ were computed
    as described in the text. $T_0$ is referred at the first, clearest
    transit event in the LC, therefore it may not coincide with the
    first transit event in the \ktwo/C5. The stellar radius $R_{\rm
      S}$ is taken from \protect\citet{Hub16}. $R_{\rm P}$ is derived
    using $R_{\rm P} / R_{\rm S}$.
  \end{tablenotes}
  \end{threeparttable}

\vspace{1cm}

  \caption{Our independent estimates of the planetary parameters of
    the two M\,44 transiting exoplanet candidates discovered by
    \protect\citet{Pope16} in \ktwo/C5 module 14.}
  \centering
  \label{Tab:exolit}
  \fontsize{7.5pt}{5pt}\selectfont
  \begin{threeparttable}
  \begin{tabular}{ccccrcccccc}
    \hline
    \hline
    & \\
    \textbf{EPIC} & \textbf{R.A.} & \textbf{Dec.} & \textbf{$K_{\rm P}$} & \multicolumn{1}{c}{\textbf{Period}} & \textbf{$T_0$} & $i$ & \textbf{$R_{\rm P} / R_{\rm S}$} & \textbf{$\delta_{\rm Phot}$} & \textbf{$R_{\rm S}$} & \textbf{$R_{\rm P}$} \\
    & [deg] & [deg] & & \multicolumn{1}{c}{[d]} & [KBJD] & [deg] & & [\%] & [$R_\odot$] & [$R_{\rm Jup}$] \\
    & \\
    \hline\hline
    & \\
    211969807 & 129.63691 & $+19.773718$ & 15.381 & $\phantom{0}1.974172 \pm 0.000089$ & $2307.382086 \pm 0.001830$ & $88.75 \pm 0.45$ & $0.0297 \pm 0.0012$ & $0.1008 \pm 0.0125$ & 0.303 & 0.088 \\
    & \\
    211990866 & 129.60134 & $+20.106105$ & 10.370 & $\phantom{0}1.673918 \pm 0.000060$ & $2341.197012 \pm 0.000776$ & $76.72 \pm 0.09$ & $0.0273 \pm 0.0007$ & $0.0621 \pm 0.0096$ & 1.572 & 0.417 \\
    & \\
    \hline
  \end{tabular}       
  \begin{tablenotes}
  \item \textbf{Notes.} See notes in Table~\protect\ref{Tab:exo}.
  \end{tablenotes}
  \end{threeparttable}   
\end{table}
\end{landscape}

%%%%%%%%%%%
\subsection{Field and M\,44 candidates description}
%%%%%%%%%%%

\begin{figure*}
  \centering
  \includegraphics[width=0.488\textwidth, keepaspectratio]{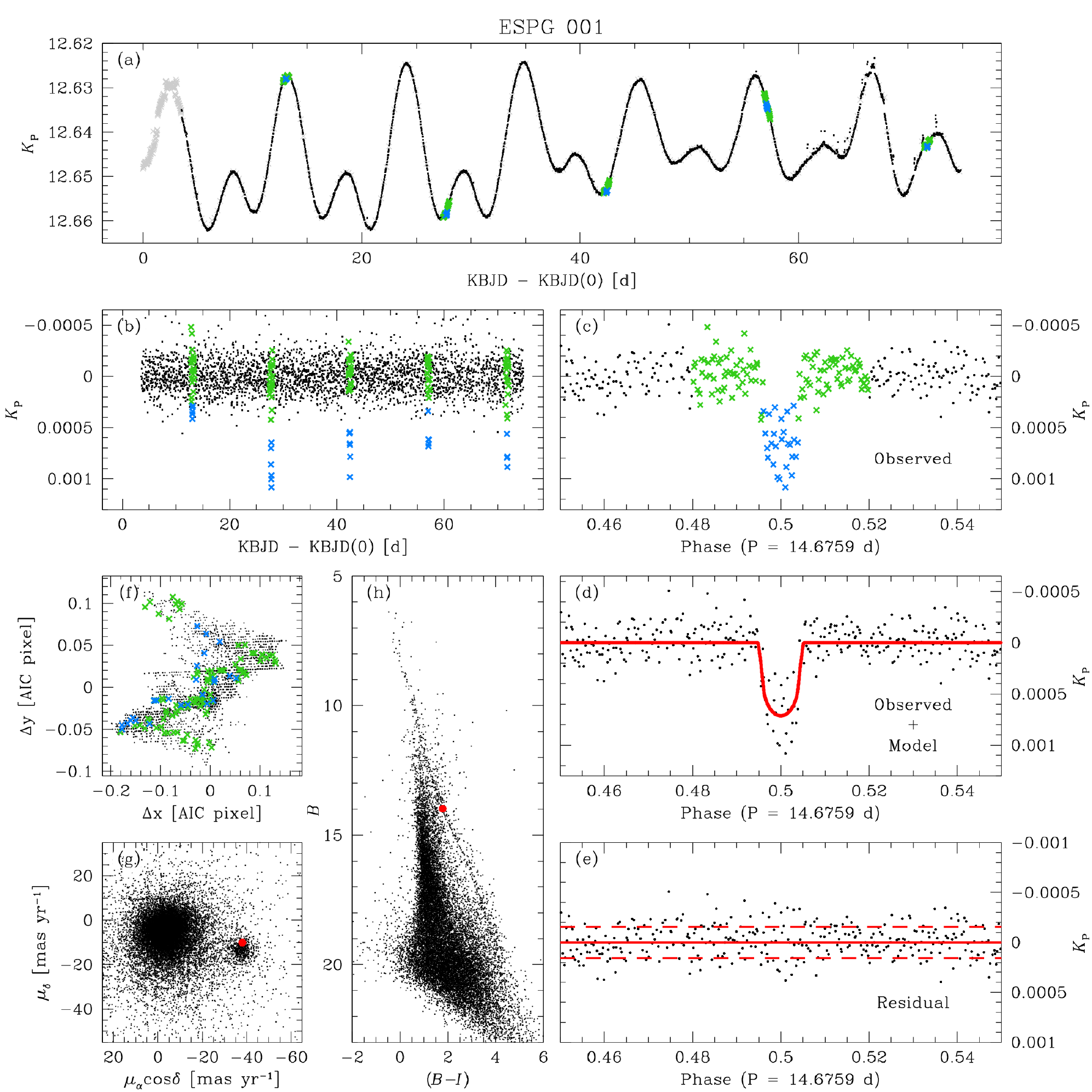}
  \includegraphics[width=0.488\textwidth, keepaspectratio]{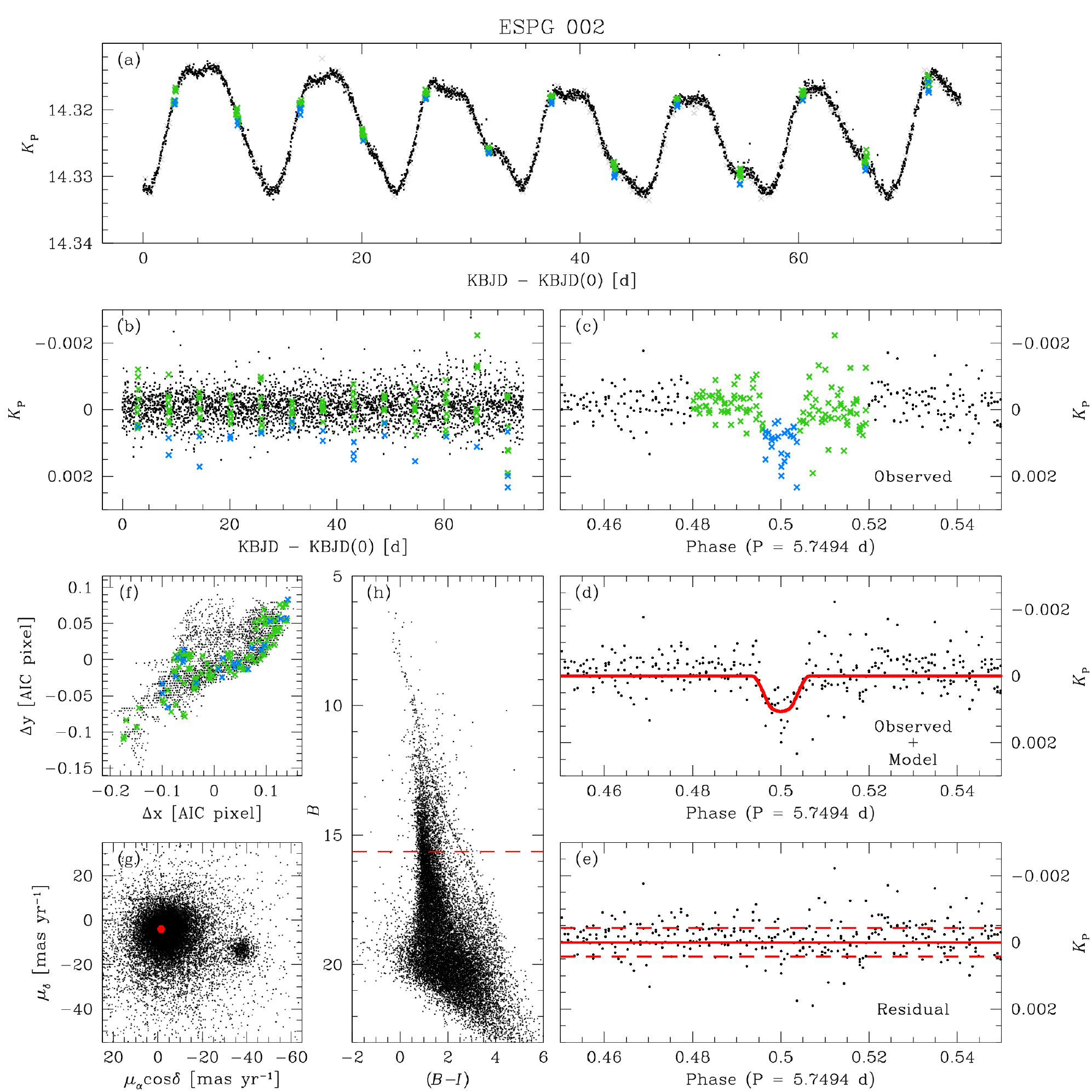}
  \caption{Summary plots for the two exoplanet candidates ESPG\,001
    and ESPG\,002. For each candidate, the detrended and the flattened
    LCs are shown in panel (a) and (b), respectively. The phased LC is
    presented in panel (c). Grey crosses represent LC points excluded
    from the analysis (e.g., thruster-jet-related events, outliers or
    noisy parts at the beginning of the LC). The centre of the transit
    is set at 0.5 phase by construction. We marked with azure crosses
    the points with $\lvert \rm Phase - 0.5 \rvert < 0.004$. These
    points roughly map the centre of the transit in the phased
    LC. Green crosses highlight the remaining transit points from
    before the ingress to after the egress of the transit ($0.004 <
    \lvert \rm Phase - 0.5 \rvert < 0.02$). In panel (d) we show again
    the phased LC and the corresponding model (red solid line)
    obtained as described in Sect.~\ref{vandm}. In panel (e) the
    difference between the observed data and the model is
    presented. The horizontal, red solid line is set at 0; while the
    horizontal, red dashed lines are set at the 68.27$^{\rm th}$
    percentile of the distribution of these residuals around the
    median. In panel (f) we show the star displacements in the
    corresponding mAIC reference-frame system. The colour-coding
    scheme is the same as that adopted in the previous panels. Note
    that the ($\Delta x$,$\Delta y$) displacements are in AIC pixels
    (1 Asiago Schmidt pixel $\sim 0.2$ \ktwo pixel). Finally, in panel
    (g) and (h) we show the vector-point diagram and the $B$
    vs. ($B-I$) CMD for the stars in the original AIC,
    respectively. The location of the exoplanet candidate is
    highlighted with a red dot. Since we do not have a $B$- and a
    $I$-filter magnitude entry for ESPG\,002 in our AIC, we drew as a
    reference a horizontal, red dashed line in the CMD using
    \textsc{EXOFOP} $B$-magnitude value.}
  \label{espg_1_2}
\end{figure*}

\begin{figure*}
  \centering
  \includegraphics[width=0.488\textwidth, keepaspectratio]{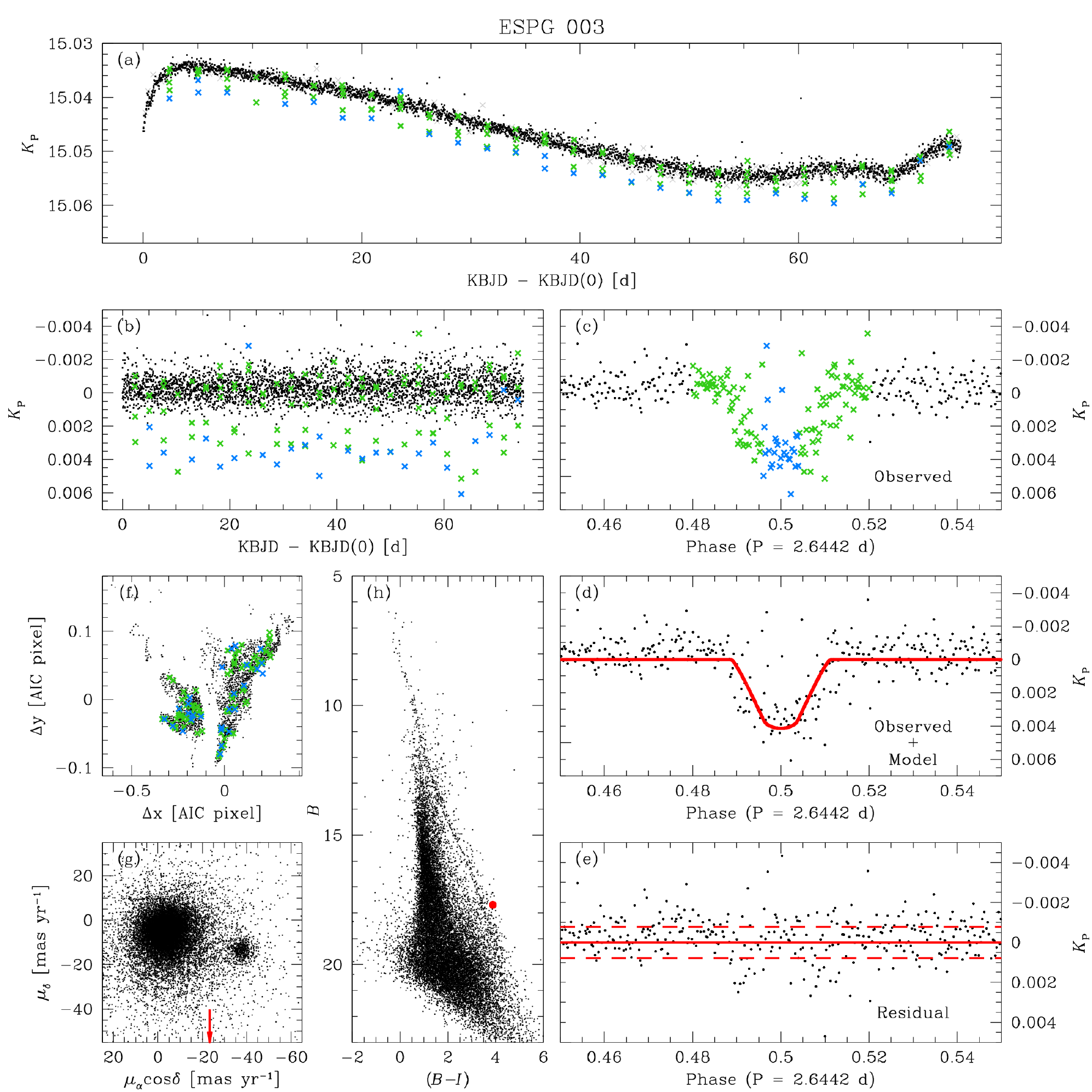}
  \includegraphics[width=0.488\textwidth, keepaspectratio]{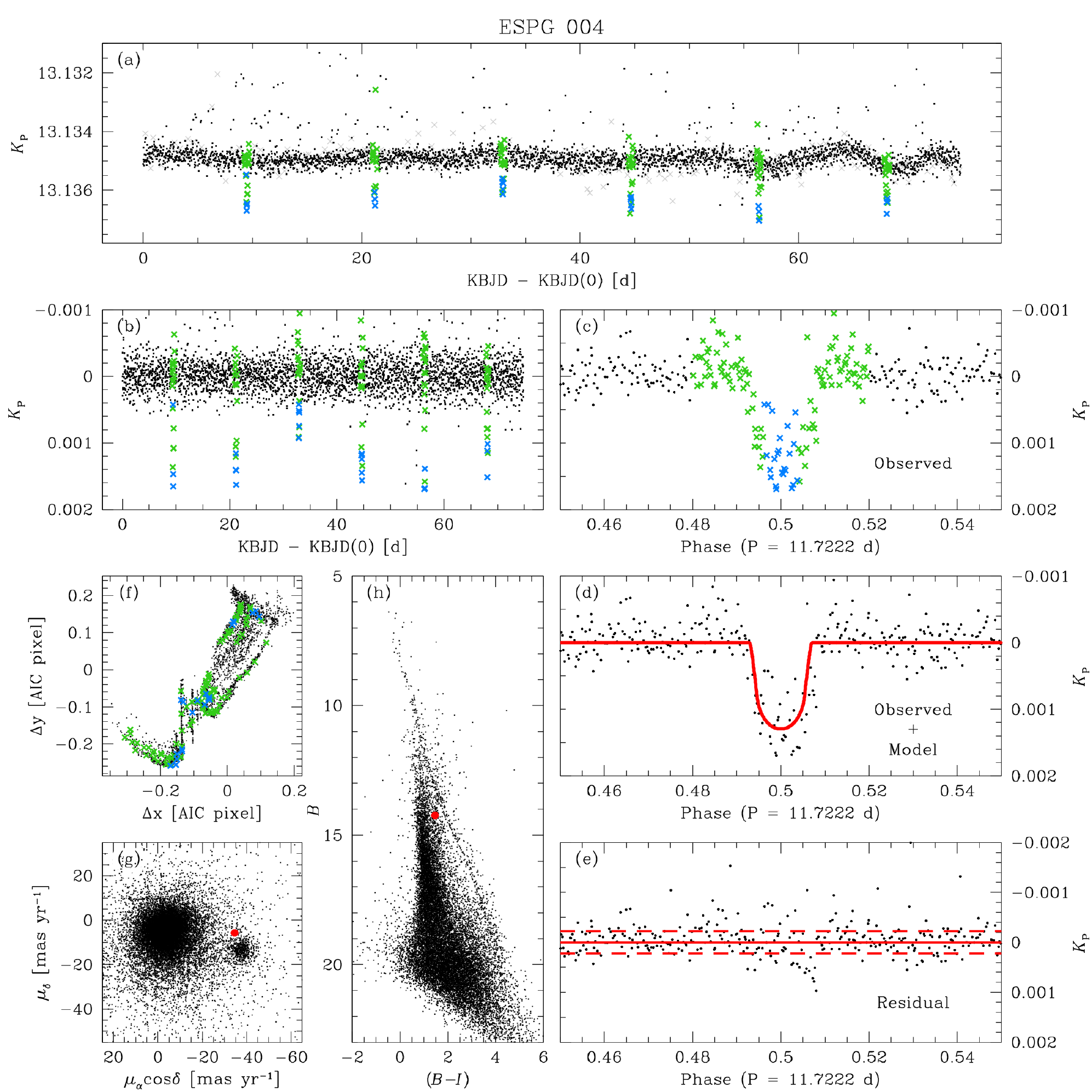}
  \caption{Same as in Fig.~\ref{espg_1_2} but for ESPG\,003 and
    ESPG\,004 candidates. For ESPG\,003, in panel (g) we marked with
    an arrow the location of the star because it lies outside the
    vector-point-diagram boundaries ($\mu_\delta = -132.6$ mas
    yr$^{-1}$).}
  \label{espg_3_4}
\end{figure*}

\begin{figure*}
  \centering
  \includegraphics[width=0.488\textwidth, keepaspectratio]{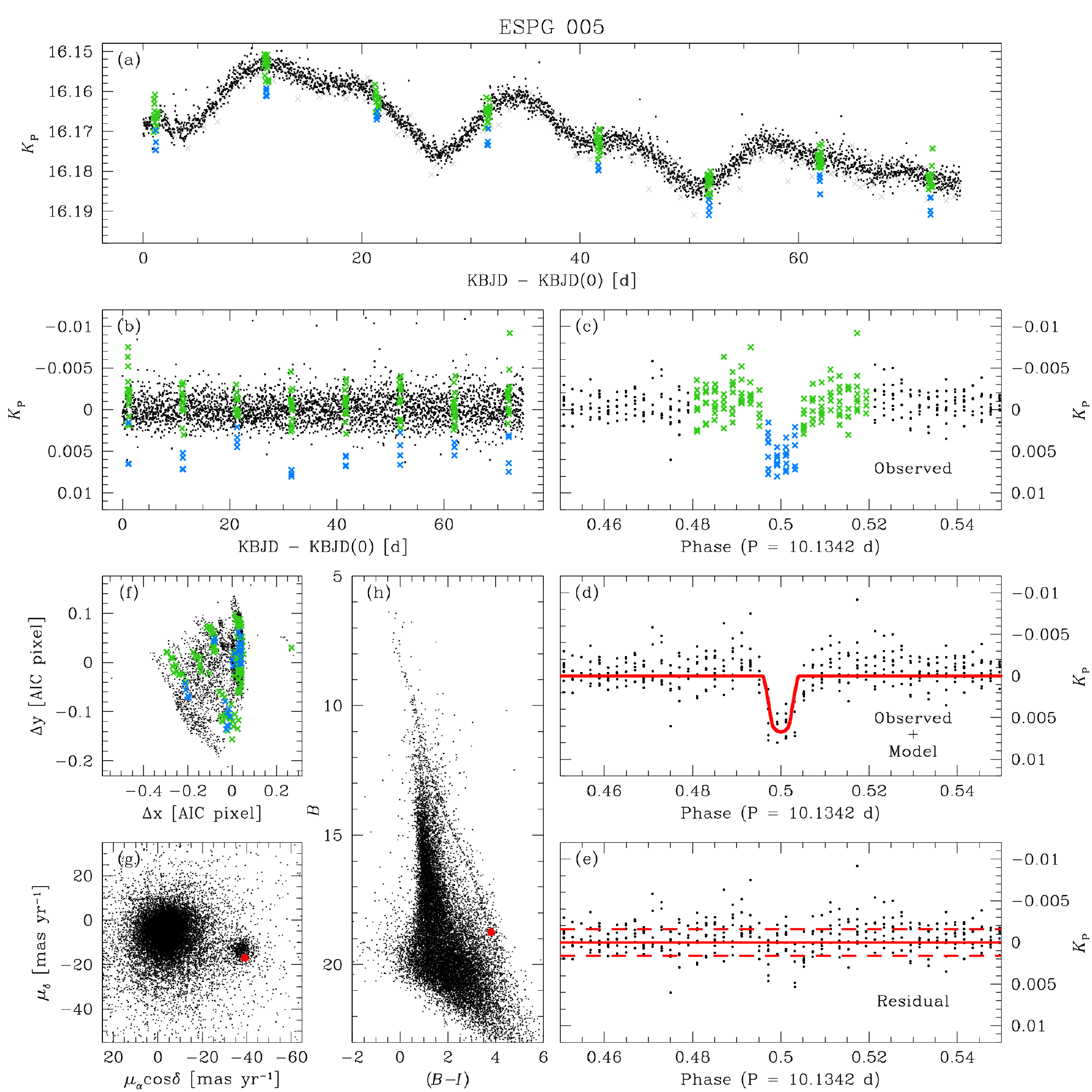}
  \includegraphics[width=0.488\textwidth, keepaspectratio]{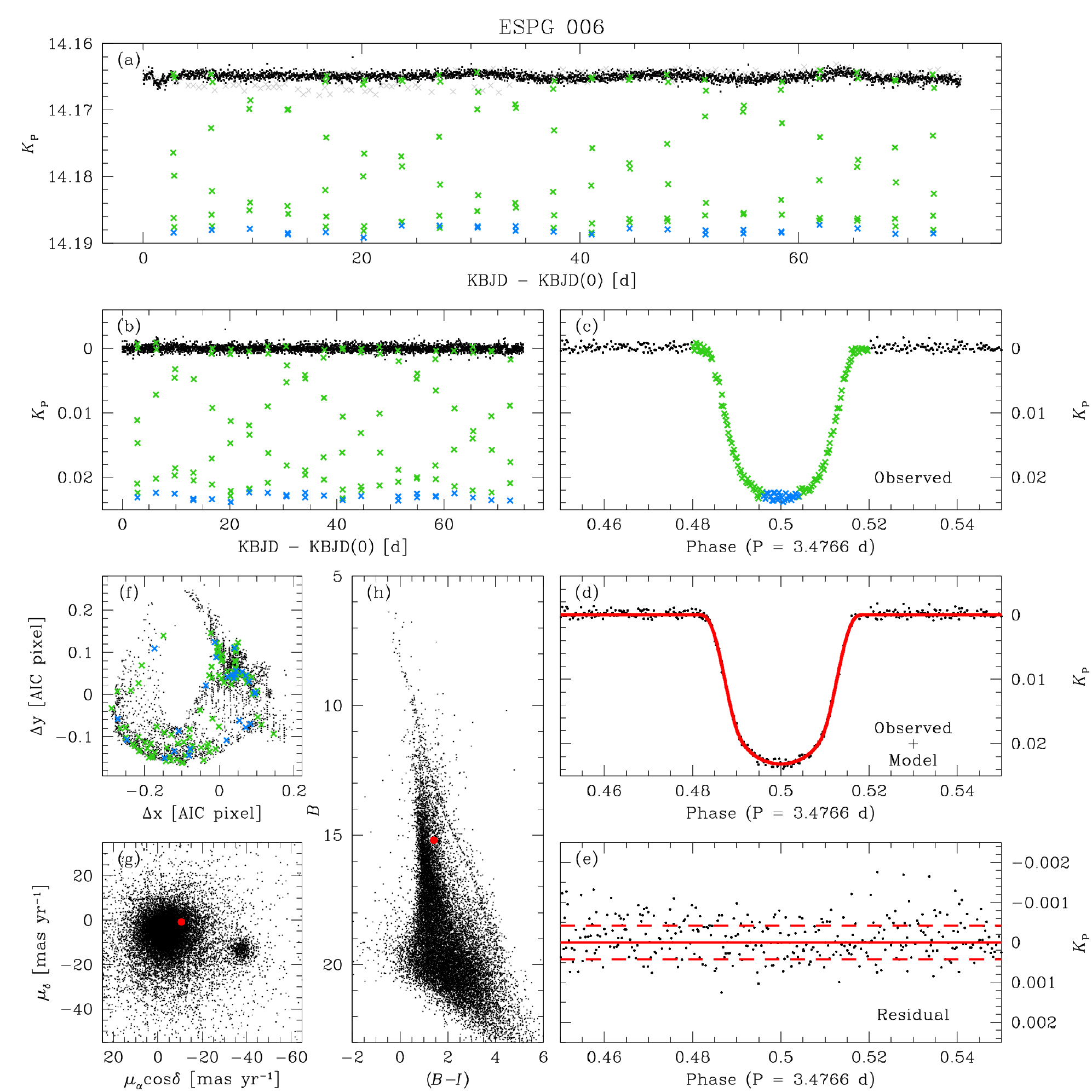}
  \caption{Same as in Fig.~\ref{espg_1_2} and \ref{espg_3_4} but for
    ESPG\,005 and ESPG\,006 candidates.}
  \label{espg_5_6}
\end{figure*}

\begin{figure*}
  \centering
  \includegraphics[width=0.488\textwidth, keepaspectratio]{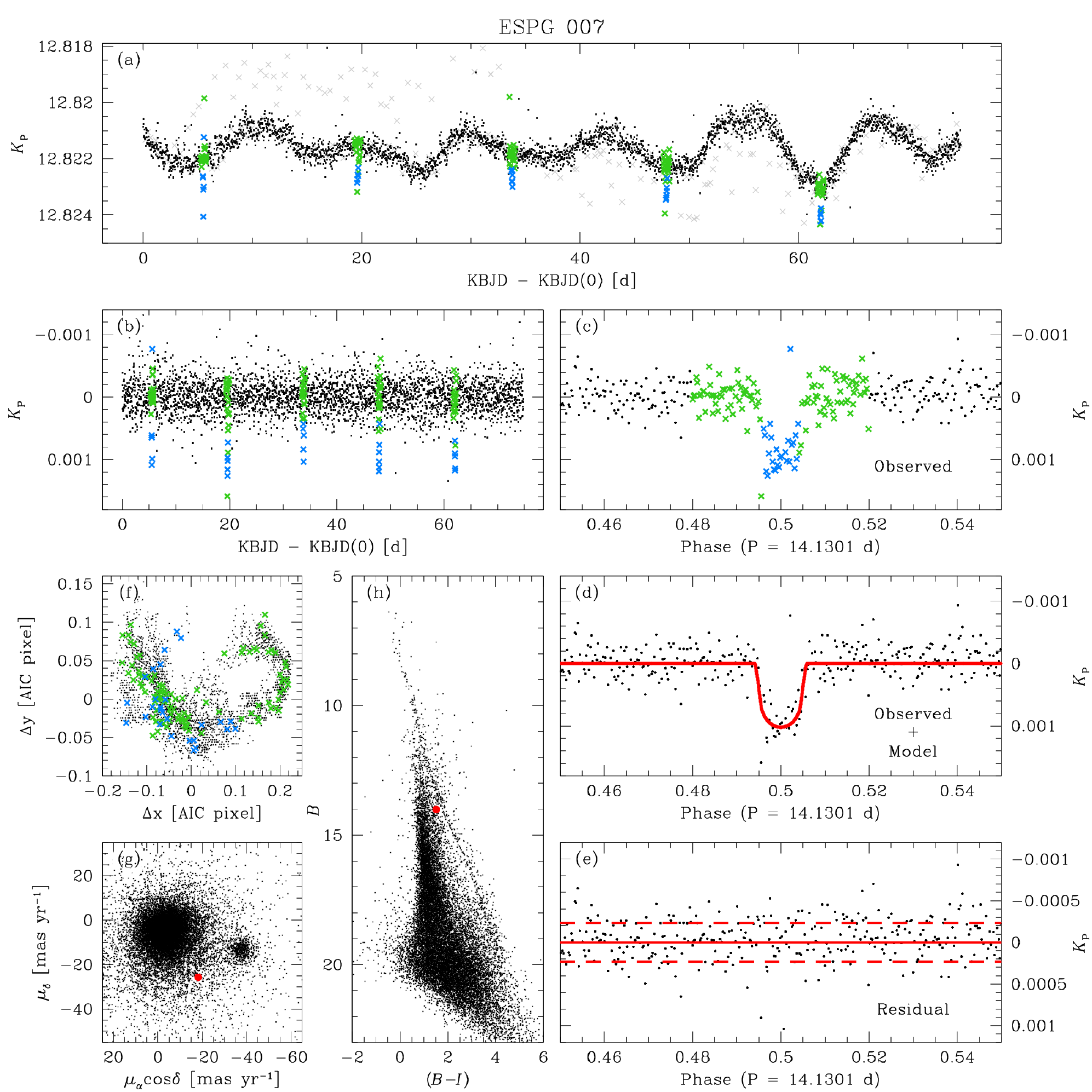}
  \caption{Same as in Fig.~\ref{espg_1_2}, \ref{espg_3_4} and
    \ref{espg_5_6} but for ESPG\,007 candidate.}
  \label{espg_7}
\end{figure*}

For each exoplanet candidate, in Fig~\ref{espg_1_2}, \ref{espg_3_4},
\ref{espg_5_6} and \ref{espg_7} we provide different plots summarising
the results. Adopting for the host stars the radius values given by
\textsc{EXOFOP}, if these signals correspond to bona-fide planets we
have detected one hot Jupiter (ESPG\,006) and six smaller (a few
$R_{\earth}$) planets.

Five out of seven objects (ESPG\,002, ESPG\,003, ESPG\,004, ESPG\,006
and ESPG\,007) are hosted by a field star, while the remaining two
candidate (ESPG\,001 and ESPG\,005) hosts are probable members of
M\,44 (on the basis of their PPMXL proper motions and CMD
locations). For these two candidate exoplanets, a RV follow-up would
be particularly important. Therefore, under simple assumptions, we can
attempt to derive indicative estimates of their expected RV signals,
and assess whether a RV follow-up is feasible or not with today's
facilities.

Planetary radii (estimated using the parameters listed in
Table~\ref{Tab:exo}) of our two candidates were converted into
indicative masses by using the probabilistic mass-radius relationship
of \citet{Wolf15} and its public-available
code\footnote{\href{https://github.com/dawolfgang/MRrelation}{https://github.com/dawolfgang/MRrelation}}. We
used the coefficients obtained from RV-based masses of planets with $R
< 4 R_{\earth}$. Confidence intervals were determined by taking the
15.865$^{\rm th}$ and the 84.135$^{\rm th}$ percentiles of the
posterior distributions, although their upper limits are set by the
maximum density allowed for a rocky planet \citep{Fort07}. We obtained
$M = (5.9 \pm 2.3) M_{\earth}$ (upper limit at 7.3 $M_{\earth}$) for
ESPG\,001 and $M = (5.6 \pm 2.3) M_{\earth}$ (upper limit at 6.9
$M_{\earth}$) for ESPG\,005. Using these planetary masses, the periods
of the planets and their inclination with respect to the line of sight
obtained from the previous LC analysis (Table~\ref{Tab:exo}), and
assuming circular orbits, we expect a RV semi amplitude of $K = ( 1.8
\pm 0.7 )$ m s$^{-1}$ (upper limit to 2.2 m s$^{-1}$) for ESPG\,001
and $K = (4.6 \pm 1.9)$ m s$^{-1}$ (upper limit to 5.6 m s$^{-1}$) for
ESPG\,005.

With the aim of obtaining an independent estimate of the host-star
masses and radii than those given by \textsc{EXOFOP}, we used our
photometry and a PARSEC (PAdova Trieste Stellar Evolution Code)
isochrone\footnote{\href{http://stev.oapd.inaf.it/cmd}{http://stev.oapd.inaf.it/cmd}}
\citep[see][and reference therein]{Bressan12,Rose16} to derive these
parameters. While for ESPG\,001 the values are in rather good
agreement, for ESPG\,005 we found a mass and a radius double than
\textsc{EXOFOP} parameters. We repeated the entire analysis for this
candidate with the new stellar mass and radius values and found a new
expected RV semi amplitude of $K = (6.1 \pm 0.3)$ m s$^{-1}$ (upper
limit to 6.8 m s$^{-1}$).

Either way, with the available facilities (e.g., \mbox{HARPS-N}@TNG),
the faintness ($V \sim 17.27$) of ESPG\,005 precludes its complete
characterisation. Therefore, the only cluster-hosted exoplanet
candidate for which a RV follow-up is possible, but challenging,
remains ESPG\,001.

The presence of spots and flares on the stellar surface are the main
responsible of the photometric modulation seen in stars of young and
intermediate-age clusters such as Praesepe. The same physical process
is affecting spectroscopic observations, and as a consequence RV
variations not due to a physical movement of the star are observed
(the so-called RV jitter). The RV jitter in M\,44 and in the
almost-coeval Hyades cluster is around 15 m s$^{-1}$
\citep{Paul04,Quinn12}. Such RV jitter could lower the sensitivity of
RV measurements to low-mass planets. However, thanks to the common
origin of RV jitter and photometric modulation, several works have
shown that when the rotational period of the star is known (from
photometry) it is possible to model and correct for the
activity-induced RV variations. This result can be achieved if a
proper observing strategy, that allows to sample both the rotational
period of the star and the period of the planets, is implemented. A
variety of successful techniques have been developed in this direction
\citep[see for example][]{Boi11,Hayw14,Faria16,Mal16}.

%%%%%%%%%%%
\subsection{Literature on exoplanets in M\,44}
%%%%%%%%%%%
\label{exolitsect}

During a RV survey focused on M\,44, \citet{Quinn12} found two hot
Jupiters and later, around one of these two stars, \citet{Mal16} also
discovered (always with RV measurements) another exoplanet. We checked
the LCs of these stars but none of them showed any transit
signature. We also checked all candidates surveyed by \citet{Quinn12}
and again found a null detection. \citet{Pep08} found two candidates
exoplanets in M\,44 field. However, accordingly to their PPMXL proper
motions, none of them is member of M\,44\footnote{For completeness, we
  also analysed the only candidate of \protect\citet{Pep08} observed
  during \ktwo/C5 (EPIC~212029841 or KP~103126). Since the target was
  imaged in channel 27, we used the public-available LCs of
  \protect\citet{Vander14} and \protect\citet{Aig16}. Phasing these
  LCs with the period given by \protect\citet{Pep08}, we did not see
  any transit-like shape. We also run our transit-search pipeline and
  obtain again a null detection. These results could mean that
  KP~103126 is not a genuine transiting exoplanet.}.

About \ktwo, \citet{Ada16} searched for ultra-short-period ($P < 1$d)
planets from Campaign~0 to 5. Only one of their candidates
(EPIC~211995325) was observed on a module-14 TPF. The object was not
detected by our pipeline because we do not search for periods shorter
than 0.5 d and the transit depth did not satisfy our selection
criteria (see Sect.~\ref{exoplanets}).

\begin{figure*}
  \centering
  \includegraphics[width=0.488\textwidth, keepaspectratio]{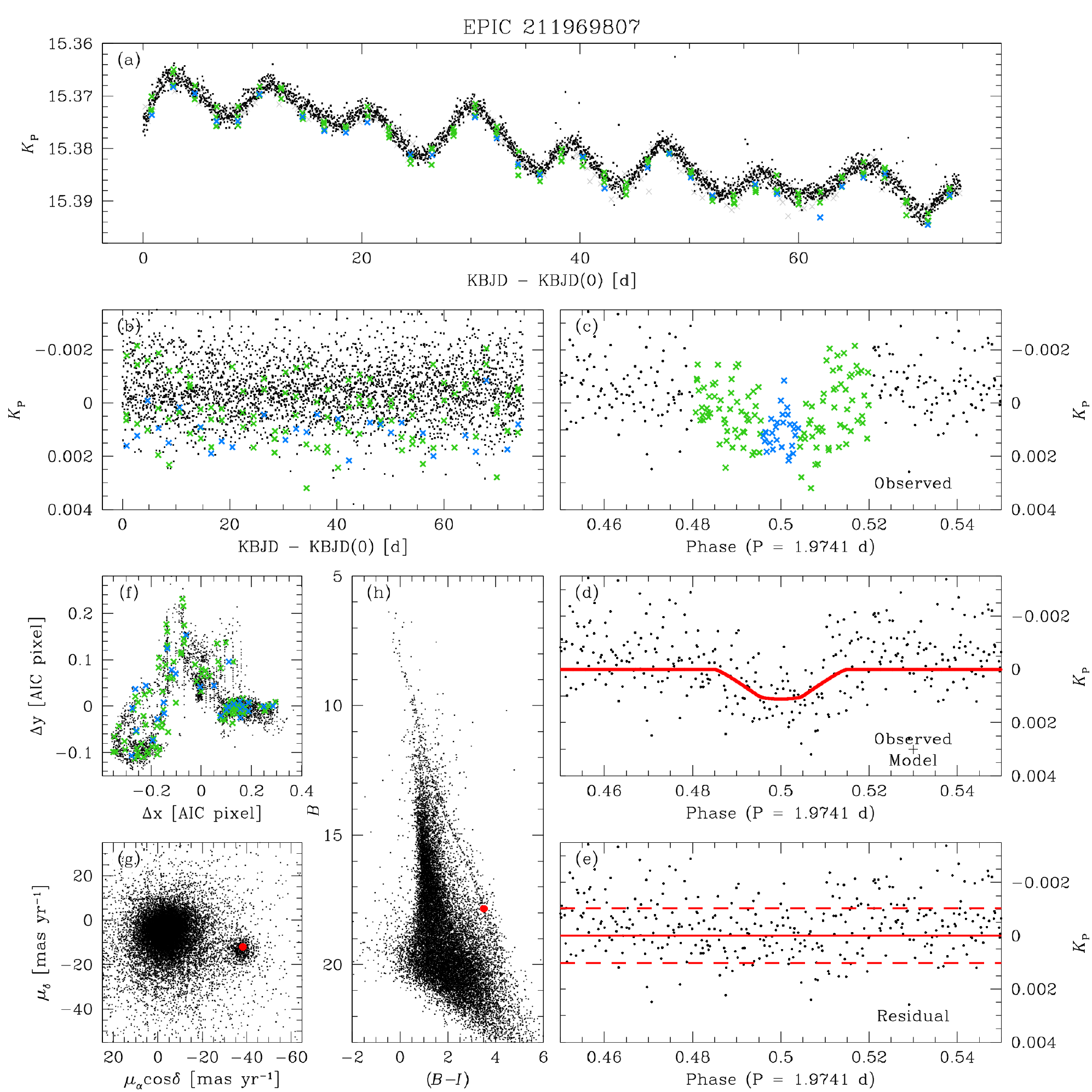}
  \includegraphics[width=0.488\textwidth, keepaspectratio]{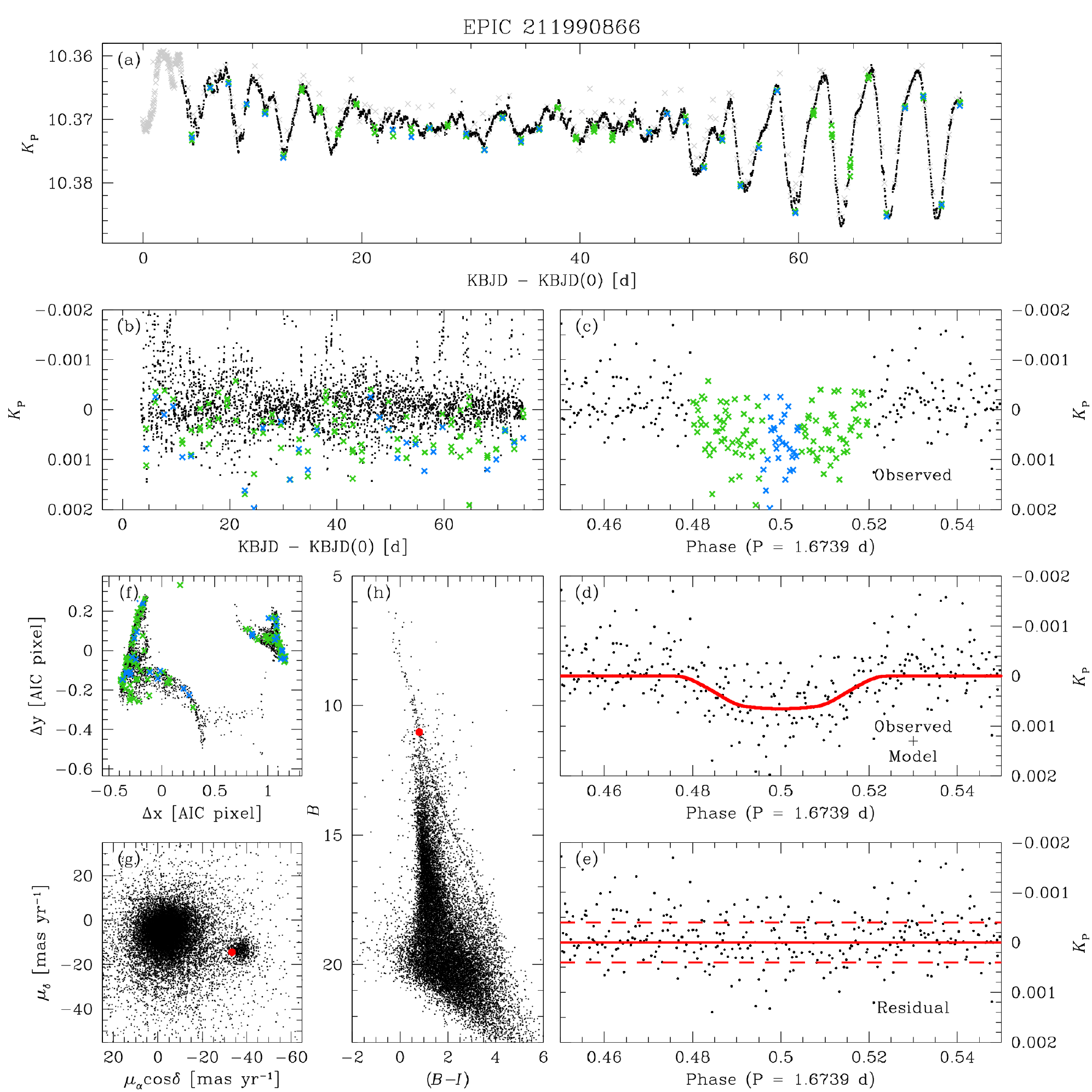}
  \caption{Summary plots for the two M\,44 exoplanet candidates
    discovered by \protect\citet{Pope16} in \ktwo/C5 module 14. See
    Fig.~\ref{espg_1_2} for a complete description of the panels. Note
    in panel (f) of EPIC~211990866 that ($\Delta x$,$\Delta y$)
    displacements can reach more than one Asiago-Schmidt pixel because
    it is saturated and its position was measured with a low
    positional accuracy.}
  \label{fig:exolit}
\end{figure*}

At the time of our submission, a work by \citet{Pope16} presenting an
independent reduction of the same \ktwo/C5 data was published. These
authors found 10 exoplanet candidates within the same \ktwo module 14
analysed in our work. Five of them were also discovered by our
pipeline (namely, ESPG\,002, ESPG\,003, ESPG\,004, ESPG\,006 and
ESPG\,007). The remaining five objects were missed because we did not
detect any significant transit signature in our LC or the transit
depth was not $1 \sigma$ below the out-of-transit level, one of the
requirements in our exoplanet finding. Accordingly to their CMD and
vector-point-diagram locations, two of such missed candidates
(EPIC~211969807 and EPIC~211990866) are M\,44 members with high
probability (see Table~\ref{Tab:exolit} and
Fig.~\ref{fig:exolit}). Note that EPIC~211990866 is saturated in \ktwo
exposures and its LC has a low photometric precision, which may
explain why we failed to identify it. We note that two of our
candidates (those hosted by M\,44 stars, ESPG\,001 and ESPG\,005) were
not detected \citet{Pope16}. This is an additional evidence that in
general every \ktwo data reduction pipeline has its pro's and con's
and still needs improvements (e.g., as done for the \kep main
mission).

Nevertheless, we will add these missing candidates (at least those
clearly visible in our LCs) to the final variable and exoplanet
catalogue we are going to release with this paper (see
Sect.~\ref{elecmat}).

\clearpage
%%%%%%%%
\section{Electronic material}
%%%%%%%%
\label{elecmat}

With this work we
release\footnote{\href{http://groups.dfa.unipd.it/ESPG/Kepler-K2.html}{http://groups.dfa.unipd.it/ESPG/Kepler-K2.html}
  and through this Journal.} all raw and detrended LCs for 1-, 1.5-,
2- and 2.5-pixel aperture and PSF photometry obtained from the
neighbour-subtracted images. We also release the \ktwo astrometrised
stacked images of the four module-14 channels.

We released a single catalogue that is the merge of the four mAICs. We
chose to merge them to simplify their usage. The catalogue is made as
follows (see Table~\ref{Tab:maiccat}). Columns (1) and (2) give the
J2000.0 equatorial coordinates in decimal degrees. Columns from (3) to
(9) provide the $N$$B$$R$$I$$J_{\rm 2MASS}$$H_{\rm 2MASS}$$K_{\rm
  2MASS}$ calibrated (except for the $N$-filter photometry)
magnitudes, when available (otherwise flagged to $-99.9999$). PPMXL
($\mu_\alpha \cos\delta$,$\mu_\delta$) proper motions are listed in
columns (10) and (11). We set the column values to $-999.99$ if the
proper motions were not available. In column (12) we give the
instrumental \kmag of the corresponding mAIC obtained as described in
Sect.~\ref{maicsub}. Finally in columns (13) and (14) we provide the
ID of the star in the corresponding mAIC and the number of the \ktwo
channel in which the star was imaged. These two columns univocally
identify the LC of the star (particularly important for the stars
added to the original AIC, see Sect.~\ref{maicsub}). For stars outside
any \ktwo channel, column (14) value was set at 0 and $K_{\rm P}$ to
$-99.9999$.

For the variable stars and exoplanet candidates we detected, we
provide to the community a catalogue with the following
columns. Column (1) gives the ID of the star in the mAIC catalogue,
while column (2) contains the channel in which the star was
imaged. Columns (3) provides the \kmag magnitude, obtained from the LC
as described in Sect.~\ref{photprec}. In column (4) we list the
variable periods, when available (e.g., for irregular or long-period
variables we set it at the \ktwo/C5 duration). Column (5) contains the
flag of our by-eye classification:
\\ $\bullet$ 1: candidate variable;
\\ $\bullet$ 2: ``difficult-interpretation'' object;
\\ $\bullet$ 3: possible blend.

Finally, in the last column (6) we give some notes about the
catalogues in the literature in which it was already described or if
the star also hosts an exoplanet candidate.

%%%%%%%%
\section{Conclusions}
%%%%%%%%

Exoplanets hosted by cluster stars are of particular interest to shed
light on the still-debated questions about their formation and
evolution. Indeed, stellar parameters (such as distance, chemistry,
mass, and age) are generally determined with a much higher accuracy
for stars in cluster rather than those in the Galactic field, giving
in return better-constrained exoplanet parameters. Furthermore,
stellar clusters are composed by an ensemble of stars with similar
properties, and such characteristic can enable a variety of
investigations, e.g, we can search for the presence of a relation
between exoplanets and host masses, understand the importance of the
dynamical evolution, or simply perform comparative analyses between
cluster stars with and without planets.

In this work we present our attempt to detect transiting exoplanet
candidates in the OC M\,44 using \ktwo/C5 data. M\,44 is one of the
few star clusters where the presence of exoplanets was firmly
confirmed with RV measurements \citep{Quinn12,Mal16}.

We applied our PSF-based techniques \citepalias[starting from the work
presented in][and improving it]{Lib16} to extract
high-photometric-precision and less-neighbour-contaminated LCs for the
stars imaged on a given module-14 TPF during \ktwo/C5. As main result
of this effort, we detected seven transiting exoplanet candidates, one
hot Jupiter and six smaller planets. Two of our candidates (ESPG\,001
and ESPG\,005) seems to be hosted by M\,44 members. Together with
those find by \citet{Pope16}, they set the number of currently-known,
transiting exoplanet candidates in M\,44 to four objects
(Table~\ref{Tab:tabover}). A RV follow-up confirmation is required to
constrain their orbital and physical parameters.

Finally, as by-product of our work, we discovered 1071 new variable
stars, tripling the number of known variables in this field to
date. Their LCs, together with those of all other objects monitored in
module 14 during \ktwo/C5, will be released via our website.

Part of our pipeline (detrending and transit search) is continuously
evolving and improving, therefore we release both raw and detrended
LCs to allow the community to not only purse their scientific goals,
but also to stimulate the development and the improvement of \ktwo
pipelines in general. This work is not a stand-alone struggle, but it
will be also very fruitful to promptly analyse the data coming from
the next exoplanet-search missions TESS \citep[Transiting Exoplanet
Survey Satellite,][]{Rick14} and PLATO \citep[PLAnetary Transits and
stellar Oscillations,][]{Rau14}.

%%%%%%%%%
\section*{Acknowledgements}
%%%%%%%%%

We acknowledge PRIN-INAF 2012 partial funding under the project
entitled ``The M4 Core Project with Hubble Space Telescope''. ML
recognizes partial support by PRIN-INAF 2014 ``The Kaleidoscope of
stellar populations in Galactic Globular Clusters with Hubble Space
Telescope''. DN and GP also acknowledge partial support by the
Universit\`a degli Studi di Padova Progetto di Ateneo CPDA141214
``Towards understanding complex star formation in Galactic globular
clusters''. LM acknowledges the financial support provided by the
European Union Seventh Framework Programme (FP7/2007-2013) under Grant
agreement number 313014 (ETAEARTH). VN acknowledges partial support by
the Universit\`a di Padova through the ``Studio preparatorio per il
PLATO Input Catalog'' grant (\#2877-4/12/15) funded by the ASI-INAF
agreement (n. 2015-019-R.0). We also thank Dr. Deokkeun An for sharing
with us its M\,44 catalogue that we used to calibrate our
Asiago-Schmidt photometry. Finally we thank the anonymous referee for
the useful comments and suggestions that improved the quality of the
paper. This research made use of the International Variable Star Index
(VSX) database, operated at AAVSO, Cambridge, Massachusetts, USA.

%%%%%%%%%%%%%%%%%%%%%%%%%%%%%%%%%%%%%%%%%%%%%%%%%%



%\bibliographystyle{mnras}


\begin{landscape}
\begin{table}
  \caption{First ten rows of the M\,44 input catalogue (merge of the four mAICs) we are going to release as electronic material.}
  \centering
  \label{Tab:maiccat}
  \fontsize{7.5pt}{5pt}\selectfont
  \begin{threeparttable}
  \begin{tabular}{ccccccccccccccc}
    \hline
    \hline
    & \\
    \textbf{Row} & \textbf{R.A.} & \textbf{Dec.} & \textbf{$N$} & \textbf{$B$} & \textbf{$R$} & \textbf{$I$} & \textbf{$J_{\rm 2MASS}$} & \textbf{$H_{\rm 2MASS}$} & \textbf{$K_{\rm 2MASS}$} & \textbf{$\mu_\alpha \cos \delta$} & \textbf{$\mu_\delta$} & \textbf{\kmag} & \textbf{ID} & \textbf{\ktwo Channel} \\
    & [deg] & [deg] & & & & & & & & [mas yr$^{-1}$] & [mas yr$^{-1}$] & & & \\
    & \\
    & (1) & (2) & (3) & (4) & (5) & (6) & (7) & (8) & (9) & (10) & (11) & (12) & (13) & (14) \\
    & \\
    \hline\hline
    & \\
    1  & 131.06866 & 20.82924 & $-14.8861$ & $-99.9999$ & $-99.9999$ & $-99.9999$ & $ 12.364$ & $ 12.045$ & $ 12.012$ & $    3.80$ & $   -9.00$ & $-12.1162$ &  1 & 48 \\
    & \\
    2  & 131.06603 & 20.76100 & $-14.5165$ & $-99.9999$ & $-99.9999$ & $-99.9999$ & $ 12.839$ & $ 12.570$ & $ 12.509$ & $    0.90$ & $   -2.30$ & $-11.7466$ &  2 & 48 \\
    & \\
    3  & 131.06539 & 20.37455 & $-10.5516$ & $-99.9999$ & $-99.9999$ & $-99.9999$ & $ 16.117$ & $ 15.296$ & $ 15.210$ & $   -0.90$ & $  -14.40$ & $ -7.7817$ &  3 & 48 \\
    & \\
    4  & 131.06347 & 20.62682 & $-13.7107$ & $-99.9999$ & $-99.9999$ & $-99.9999$ & $ 13.098$ & $ 12.654$ & $ 12.504$ & $   -8.70$ & $   -3.50$ & $-10.9408$ &  4 & 48 \\
    & \\
    5  & 131.06191 & 20.12612 & $-14.1353$ & $-99.9999$ & $-99.9999$ & $-99.9999$ & $ 12.426$ & $ 11.892$ & $ 11.761$ & $  -14.30$ & $  -23.80$ & $-11.3654$ &  5 & 48 \\
    & \\
    6  & 131.06303 & 20.98408 & $-13.0740$ & $-99.9999$ & $-99.9999$ & $-99.9999$ & $ 13.725$ & $ 13.274$ & $ 13.222$ & $   -7.60$ & $  -13.00$ & $-99.9999$ &  6 &  0 \\
    & \\
    7  & 131.06130 & 20.02970 & $-12.1400$ & $-99.9999$ & $-99.9999$ & $-99.9999$ & $ 14.847$ & $ 14.548$ & $ 14.513$ & $    0.00$ & $   -9.60$ & $ -9.3701$ &  7 & 48 \\
    & \\
    8  & 131.06210 & 20.65114 & $-13.1195$ & $-99.9999$ & $-99.9999$ & $-99.9999$ & $ 13.990$ & $ 13.665$ & $ 13.555$ & $   -7.10$ & $   -5.50$ & $-10.3496$ &  8 & 48 \\
    & \\
    9  & 131.06153 & 20.62936 & $-10.6903$ & $-99.9999$ & $-99.9999$ & $-99.9999$ & $ 16.324$ & $ 16.115$ & $ 15.661$ & $   -3.00$ & $   -3.40$ & $ -7.9204$ &  9 & 48 \\
    & \\
    10 & 131.05881 & 18.94697 & $-13.5582$ & $-99.9999$ & $-99.9999$ & $-99.9999$ & $-99.9999$ & $-99.9999$ & $-99.9999$ & $ -999.99$ & $ -999.99$ & $-10.8129$ & 10 & 45 \\
    & \\
    (...) & (...) & (...) & (...) & (...) & (...) & (...) & (...) & (...) & (...) & (...) & (...) & (...) & (...) & (...) \\
    & \\
    \hline
  \end{tabular}
  \end{threeparttable}  

  \vspace{1 cm}
  
  \caption{First ten rows of the variable/exoplanet catalogue.}
    \centering
  \label{Tab:maiccat}
  \fontsize{7.5pt}{5pt}\selectfont
  \begin{threeparttable}
  \begin{tabular}{ccccccc}
    \hline
    \hline
    & \\
    \textbf{Row} & \textbf{ID} & \textbf{\ktwo Channel} & \textbf{LC \kmag} & \textbf{$P$} & \textbf{Flag} & \textbf{Notes} \\
    & & & & [d] & & \\
    & \\
    & \\
    & (1) & (2) & (3) & (4) & (5) & (6) \\
    & \\
    \hline\hline
    & \\
    1  &  17 & 45 & 12.35 & 11.09064293 & 1 & \\
    2  &  29 & 45 & 14.32 & 14.21546437 & 1 & \\
    3  &  30 & 45 & 13.61 & 12.19786185 & 1 & 6, 8 \\
    4  &  45 & 45 & 12.21 &  9.46942449 & 1 & \\
    5  &  59 & 45 & 15.65 & 74.82000000 & 1 & \\
    6  & 119 & 45 & 13.52 & 74.82000000 & 1 & \\
    7  & 267 & 45 & 12.05 & 74.82000000 & 1 & 12 \\
    8  & 298 & 45 & 13.70 & 74.82000000 & 1 & \\
    9  & 304 & 45 & 16.76 & 10.42804406 & 2 & \\
    10 & 307 & 45 & 16.48 & 18.14582914 & 2 & \\

    & \\
    (...) & (...) & (...) & (...) & (...) & (...) & (...) \\
    & \\
    \hline
  \end{tabular}       
  \begin{tablenotes}
  \item \textbf{Notes.} Column (6) lists the catalogues in literature
    that already analysed the object (1 = \protect\citet{Agu11}, 2 =
    \protect\citet{Bou01}, 3 = \protect\citet{Bre12}, 4 =
    \protect\citet{Cas12}, 5 = \protect\citet{Del11}, 6 =
    \protect\citet{Dou14}, 7 = \protect\citet{Dra14}, 8 =
    \protect\citet{Kov14}, 9 = \protect\citet{Li07}, 10 =
    \protect\citet{Liu07}, 11 = \protect\citet{Mal16}, 12 =
    \protect\citet{Mer09}, 13 = \protect\citet{Pep08}, 14 =
    \protect\citet{Quinn12}, 15 = \protect\citet{Sch11}, 16 = GCVS, 17
    = VSX) and/or states if the star hosts one of the candidate
    exoplanets listed in Tables~\ref{Tab:exo} and \ref{Tab:exolit}.
  \end{tablenotes}
  \end{threeparttable}  

\end{table}
\end{landscape}
\begin{table}
  \caption{List of M\,44 candidate and confirmed exoplanets.}
  \centering
  \label{Tab:tabover}
  \fontsize{7.5pt}{5pt}\selectfont
  \begin{threeparttable}
  \begin{tabular}{cccc}
    \hline
    \hline
    & \\
    \textbf{EPIC} & \textbf{R.A.} & \textbf{Dec.} & \textbf{Notes}\\
    & [deg] & [deg] & \\
    & \\
    \hline\hline
    & \\
    & \\
    \multicolumn{4}{c}{\textit{Candidates}} \\
    & \\
    & \\
    211913977 & 130.34349 & $+18.934026$ & ESPG\,001 \\
    & \\
    211916756 & 129.36243 & $+18.976653$ & ESPG\,005 \\
    & \\    
    211969807 & 129.63691 & $+19.773718$ & \\
    & \\
    211990866 & 129.60134 & $+20.106105$ & \\
    & \\
    & \\
    \multicolumn{4}{c}{\textit{Confirmed}} \\
    & \\
    & \\
    211998346 & 130.43243 & $+20.226899$ & Pr201b \\
    & \\
    211936827 & 130.54745 & $+19.277061$ & Pr211b,c \\
    & \\
    \hline
  \end{tabular}    
  \begin{tablenotes}
  \item \textbf{Notes.} EPIC 211998346 and 211936827 are the two M\,44
    stars that host the three RV-confirmed exoplanets discovered by
    \protect\citet{Quinn12} and \protect\citet{Mal16}.
  \end{tablenotes}
  \end{threeparttable}   
\end{table}

%%%%%%%%%%%%%%%%%%%%%%%%%%%%%%%%%%%%%%%%%%%%%%%%%%

% Don't change these lines
\bsp	% typesetting comment
\label{lastpage}
\end{document}